\documentclass[fleqn,usenatbib]{mnras}

\usepackage{newtxtext,newtxmath}

\usepackage[T1]{fontenc}

\DeclareRobustCommand{\VAN}[3]{#2}
\let\VANthebibliography\thebibliography
\def\thebibliography{\DeclareRobustCommand{\VAN}[3]{##3}\VANthebibliography}

\usepackage{graphicx}	
\usepackage{amsmath}	
\usepackage{bm}		
\usepackage{xcolor}

\newcommand{\Msun}{\text{M}_{\odot}}   
\newcommand{\MsunPerYr}{\text{M}_{\odot} \, \text{yr}^{-1}}   
\newcommand{\Order}[1]{\mathcal{O}(#1)}
\newcommand{\sgn}{\text{sgn}}

\title[GW from r-mode of stochastically accreting NS]{Gravitational waves from r-mode oscillations of stochastically accreting neutron stars}

\author[W. Dong \& A. Melatos]{
Wenhao Dong,$^{1,2}$\thanks{E-mail: wddong@student.unimelb.edu.au}
Andrew Melatos,$^{1,2}$\thanks{E-mail: amelatos@unimelb.edu.au}
\\
$^{1}$School of Physics, University of Melbourne, Parkville, VIC 3010, Australia. \\
$^{2}$ARC Centre of Excellence for Gravitational Wave Discovery (OzGrav), University of Melbourne, Parkville, VIC 3010, Australia.
}

\date{Accepted XXX. Received YYY; in original form ZZZ}

\pubyear{\the\year{}}

\begin{document}
\label{firstpage}
\pagerange{\pageref{firstpage}--\pageref{lastpage}}
\maketitle

\begin{abstract}
$r$-mode oscillations in rotating neutron stars are a source of continuous gravitational radiation.
We investigate the excitation of $r$-modes by the mechanical impact on the neutron star surface of stochastically accreted clumps of matter, assuming that the Chandrasekhar-Friedman-Schutz instability is not triggered.
The star is idealised as a slowly-rotating, unmagnetised, one-component fluid with a barotropic equation of state in Newtonian gravity.
It is found that the $r$-mode amplitude depends weakly on the equation of state but sensitively on the rotation frequency \(\nu_{\rm s}\). The gravitational wave strain implicitly depends on the equation of state through the damping timescale.
The root-mean-square strain is \(h_{\rm rms} \approx 10^{-35} (\nu_{\rm s}/ 10\,{\rm Hz})^{2} (R_*/10\,{\rm km})^2 (\Delta t_{\rm acc}/1\,{\rm yr})^{1/2} (f_{\rm acc}/1\,{\rm kHz})^{-1/2} (\dot{M}/10^{-8}\MsunPerYr) (v/0.4c) (d/1\,{\rm kpc})^{-1}\), which is comparable to the strain from $g$-, $p$- and $f$-modes excited by stochastic accretion, where \(R_*\) is the radius of the star, \(\Delta t_{\rm acc}\) is the uninterrupted duration of an accretion episode, \(f_{\rm acc}\) is the mean clump impact frequency, \(\dot{M}\) is the accretion rate, \(v\) is the impact speed, and \(d\) is the distance of the star from the Earth.
An observational test is proposed, based on the temporal autocorrelation function of the gravitational wave signal, to discern whether the Chandrasekhar-Friedman-Schutz instability switches on and coexists with impact-excited $r$-modes before or during a gravitational wave observation.
\end{abstract}

\begin{keywords}
accretion, accretion discs -- asteroseismology -- gravitational waves -- stars: neutron -- stars: oscillations -- stars: rotation
\end{keywords}



\section{Introduction}

$r$-modes are inertial modes in rotating neutron stars, whose restoring force is the Coriolis force.
Gravitational waves (GWs) generated by $r$-mode oscillations have attracted significant attention in continuous wave searches \citep{Riles2023,Wette2023}.
This is because $r$-modes grow unstably through the emission of gravitational radiation \citep{Andersson1998,FriedmanMorsink1998,LindblomEtAl1998,OwenEtAl1998}.
Recent targeted and all-sky searches for GWs from $r$-modes report no detections and infer upper bounds on the characteristic wave strain satisfying \(h_{0}^{95\%} \lesssim 10^{-25}\) at \(95\%\) confidence, using data from the first three observing runs of the Laser Interferometer Gravitational-wave Observatory (LIGO), Virgo, and Kamioka Gravitational Wave Detector (KAGRA) \citep{FesikPapa2020,MiddletonEtAl2020,AbbottEtAl2021,AbbottEtAl2021b,RajbhandariEtAl2021,CovasEtAl2022,LIGOScientificCollaborationEtAl2022,LIGOScientificCollaborationAndVirgoCollaborationEtAl2022,VargasMelatos2023}.
Indirect upper limits on \(h_0^{95\%}\) from $r$-modes can also be inferred from electromagnetic observations of neutron stars, assuming that the gravitational radiation reaction torque causes the star to spin down \citep{Owen2010,GlampedakisGualtieri2018,CarideEtAl2019,MiddletonEtAl2020} or balances the accretion torque \citep{Wagoner1984,Bildsten1998,GlampedakisGualtieri2018,MiddletonEtAl2020}, or that photon cooling is in thermal equilibrium with the viscous damping of $r$-modes \citep{SchwenzerEtAl2017}.

Unstable $r$-modes grow exponentially from small, random, initial fluctuations, until nonlinear mode coupling saturates their growth \citep{SchenkEtAl2001,ArrasEtAl2003,BondarescuEtAl2007}.
The initial fluctuations are sometimes attributed to nonspecific origins (e.g.\ thermal) and sometimes modelled in detail.
For example, buoyant $r$-modes may be excited in the surface ocean during thermonuclear type I X-ray bursts \citep{StrohmayerLee1996,Heyl2004,ChambersWatts2020} 
and can lead to coherent X-ray oscillations during a superburst, as claimed in XTE J1751$-$305 \citep{StrohmayerMahmoodifar2014}.
The excited surface $r$-mode is not related directly to the $r$-modes in the bulk of the star, which themselves may be amplified near the surface \citep{Lee2014}.
In binary systems accreting at present, oscillation modes can also be excited by repeated mechanical impacts of clumps of stochastically accreted matter \citep{NagarEtAl2004,DongMelatos2024}. 
Previous work along these lines has focused on nonradial $f$-, $p$-, and $g$-modes in a non-rotating star with a polytropic equation of state (EOS). 
The root-mean-square GW strain increases with increasing polytropic index, attaining \(10^{-33} \leq h_{\rm rms} \leq 10^{-32}\) under astrophysically plausible conditions \citep{DongMelatos2024}.

In this paper, we extend previous calculations to include rotation and investigate nonradial $r$-mode oscillations excited mechanically by stochastic accretion. 
We calculate the amplitude of the excited $r$-modes and the root-mean-square amplitude and Fourier spectrum of the resulting gravitational radiation, assuming that the star rotates slowly enough not to trigger the Chandrasekhar-Friedman-Schutz (CFS) instability.
The paper is organised as follows. 
In Section~\ref{sec:2-r-mode_oscillations_in_slowly_rotating_stars}, we introduce the stellar model and briefly review the linear perturbation theory of free $r$-mode oscillations.
In Section~\ref{sec:3-Accretion_excitation}, we calculate the impulsive $r$-mode response to the mechanical impact of accreted matter, first from a single clump and then from a stochastic sequence of clumps, as implied by modern numerical simulations \citep{RomanovaOwocki2015}.
We compute the GW signal from the excited $r$-modes as a function of the ensemble statistics of the accretion clumps in Section~\ref{sec:4-gravitational_radiation}.
We review the conditions for exciting the CFS (in)stability and discuss the astrophysical implications of the cyclically triggered CFS instability in Section~\ref{sec:5-CFS_instability}.
Section~\ref{sec:6-conclusions} is the conclusion.

\section{r-mode oscillations in a slowly rotating star}
\label{sec:2-r-mode_oscillations_in_slowly_rotating_stars}

The neutron star model in this paper is kept as simple as possible to focus on the new feature of the analysis: stochastic excitation of $r$-modes by the mechanical impacts of clumps of accreting matter.
We model the star as a one-component ideal fluid rotating uniformly with angular velocity \(\bm{\Omega}\) in Newtonian gravity. 
We also assume that the EOS is barotropic.
Equilibria and $r$-mode oscillations of this idealised model have been examined in detail in the literature \citep{Tassoul1978,PapaloizouPringle1978,Saio1982}, as well as their generalised counterparts in the broader inertial modes family \citep{LindblomIpser1999,LockitchFriedman1999}.
We neglect the corrections from rapid rotation \citep{LindblomEtAl1999}, shear viscosity at the crust-core boundary \citep{BildstenUshomirsky1999,LevinUshomirsky2001}, stratification \citep{YoshidaLee2000,PassamontiEtAl2009,AnderssonGittins2023,GittinsAndersson2023}, nonlinear mode coupling \citep{SchenkEtAl2001,ArrasEtAl2003,BondarescuEtAl2007}, superfluidity \citep{LindblomMendell2000,AnderssonComer2001}, and magnetic fields \citep{RezzollaEtAl2000,MorsinkRezania2002}.
Although pure $r$-modes do not exist in general relativistic barotropes, relativistic corrections to the Newtonian $r$-modes have also been calculated using axial-led inertial modes in relativistic barotropic stars \citep{LockitchEtAl2000,LockitchEtAl2003,IdrisyEtAl2015,GhoshEtAl2023a}.
For relativistic nonbarotropic stars, however, the eigenvalue problem of pure $r$-modes is singular \citep{Kojima1998}.
These and other realistic corrections can be incorporated into the theoretical framework Sections~\ref{sec:2-r-mode_oscillations_in_slowly_rotating_stars} and \ref{sec:3-Accretion_excitation}, when the need arises. They are essential for a detailed analysis of GW observations, once detections are made. 

In this section, we review the linear perturbation theory of $r$-mode oscillations in a slowly rotating star.
In Sections~\ref{subsec2.1:Equilibrium} and \ref{subsec2.2:Linearised equations of motion}, we set out the equilibrium configuration and the linearised equations of motion, respectively. The $r$-mode oscillations are assumed to be adiabatic. We introduce the slow-rotation expansion and derive the eigenfrequencies and eigenfunctions for $r$-modes in Section~\ref{subsec2.3:Slow_rotation_expansion}. Orthogonality of the eigenmodes is discussed in Section~\ref{subsec2.4:Orthogonality}.

\subsection{Equilibrium}
\label{subsec2.1:Equilibrium}
The mass continuity, Euler, and Poisson equations in the corotating frame are
\begin{align}
    &\frac{\partial \rho}{\partial t} + \nabla \cdot (\rho \bm{u}) 
    = 0,
    \label{eqn:mass continuity rot_frame}
    \\
    &\frac{\partial \bm{u}}{\partial t} + (\bm{u} \cdot \nabla) \bm{u} + 2 \bm{\Omega} \times \bm{u} + \bm{\Omega} \times (\bm{\Omega} \times \bm{x})
    = -\frac{1}{\rho} \nabla P - \nabla \Phi ,
    \label{eqn:Euler equation rot_frame}
    \\
    &\nabla^2 \Phi 
    = 4\pi G \rho,
    \label{eqn:Poisson equation rot_frame}
\end{align}
where \(\bm{u} = \bm{v} - \bm{\Omega} \times \bm{x}\) is the flow velocity in the corotating frame, \(\bm{v}\) is the flow velocity in the inertial frame, \(\bm{x}\) is the position vector, \(\rho\) is the density, \(P\) is the pressure, and \(\Phi\) is the gravitational potential.

At equilibrium, the star obeys the unperturbed stationary Euler equation with \(\bm{u}_0 = 0\), viz.
\begin{equation}
    0 = -\frac{1}{\rho_0} \nabla P_0 - \nabla \Phi_0 - \nabla \Phi_{\text{rot}, 0} ,
    \label{eqn:Euler equation background}
\end{equation}
where \(\Phi_{\rm rot} = - (\bm{\Omega}\times\bm{x})^2/2\) is the centrifugal potential. The subscript zero denotes an equilibrium quantity.
For slowly rotating stars satisfying \(|\bm{\Omega}| \ll (\pi G \rho_0)^{1/2}\), the centrifugal potential is negligible compared to the gravitational potential, and the background star is approximately spherically symmetric.

\subsection{Linearised equations of motion}
\label{subsec2.2:Linearised equations of motion}
We describe the linear perturbations of the star by Eulerian perturbed quantities, which are denoted by the operator prefix \(\delta\), e.g.\ the pressure perturbation is \(\delta P\).
Lagrangian perturbations, denoted by the prefix \(\Delta\), are related to Eulerian perturbations via \(\bm{\xi}(\bm{x}, t)\), the Lagrangian displacement of a fluid element from its equilibrium position. Mathematically, we write
\begin{align}
    \Delta = \delta +  \bm{\xi} \cdot \nabla .
    \label{eqn:Lagrangian Eulerian relation}
\end{align}
With \(\bm{u}_0 = 0\) and \(\delta \bm{u} = \Delta \bm{u}\), we obtain
\begin{align}
    \Delta \bm{u} = \frac{\partial \bm{\xi}}{\partial t} .
\end{align}
The linearised forms of equations~\eqref{eqn:mass continuity rot_frame}--\eqref{eqn:Poisson equation rot_frame}, expressed in terms of Eulerian perturbations, are then given by 
\begin{align}
    &0 = \frac{\partial \delta\rho}{\partial t} + \nabla \cdot \left(\rho_0 \frac{\partial \bm{\xi}}{\partial t}\right) , 
    \label{eqn:mass continuity perturbed}
    \\
    &0 = \frac{\partial^2 \bm{\xi}}{\partial t^2} + \mathbfss{B} \cdot \frac{\partial \bm{\xi}}{\partial t} + \mathbfss{C} \cdot \bm{\xi} ,
    \label{eqn:Euler equation perturbed}
    \\
    &4\pi G \delta\rho
    = \nabla^2 \delta\Phi ,
    \label{eqn:Poisson equation perturbed}
\end{align}
where \(\mathbfss{B}\) has Cartesian components
\begin{align}
    B_{ik} &= 2 \Omega_j \epsilon_{ijk} ,
    \label{eqn:operator B}
\end{align}
\(\epsilon_{ijk}\) is the Levi-Civita symbol, the Einstein summation convention applies to repeated indices, and \(\mathbfss{C}\) satisfies \citep{Lynden-BellOstriker1967,SchenkEtAl2001}
\begin{align}
    \mathbfss{C} \cdot \bm{\xi} &= -\frac{\nabla \delta P}{\rho_0} + \frac{\delta\rho}{\rho_0^2} \nabla P_0 - \nabla \delta\Phi .
    \label{eqn:operator C}
\end{align}
Note that the centrifugal term in equation~\eqref{eqn:Euler equation rot_frame} does not appear in equation~\eqref{eqn:Euler equation perturbed}, because \(\Delta[\bm{\Omega} \times (\bm{\Omega} \times \bm{x})]\) cancels with \((\bm{\xi} \cdot \nabla)(-\nabla P_0 / \rho_0 - \nabla \Phi_0)\) on the right-hand side of equation~\eqref{eqn:Euler equation rot_frame}.

We assume that the perturbations are adiabatic, satisfying
\begin{align}
    \frac{\Delta P}{P_0} = \Gamma_1 \frac{\Delta \rho}{\rho_0} ,
    \label{eqn:adiabatic condition}
\end{align}
where \(\Gamma_1\) is the adiabatic index. In general, \(\Gamma_1 = \Gamma_1(\bm{x})\) is a function of the background structure of the star.
Equation~\eqref{eqn:Euler equation perturbed} together with equation~\eqref{eqn:adiabatic condition} reduces to equation~(14) in \citet{DongMelatos2024} for \(\bm{\Omega}=0\).
With equation~\eqref{eqn:Lagrangian Eulerian relation}, equation~\eqref{eqn:adiabatic condition} can also be written as
\begin{align}
    \frac{\delta P}{P_0} = \Gamma_1 \frac{\delta \rho}{\rho_0} ,
    \label{eqn:adiabatic condition 2}
\end{align}
under the assumption that the Schwarzschild discriminant \(\bm{A}\) satisfies
\begin{align}
    \bm{A} &= \frac{1}{\rho_0} \nabla \rho_0 - \frac{1}{\Gamma_1 P_0} \nabla P_0 
    \\
    &= 0.
    \label{eqn:Schwarzschild discriminant condition}
\end{align}
Equation~\eqref{eqn:Schwarzschild discriminant condition} holds for barotropic stars, where \(P\) is a function of \(\rho\) only [\(P = P(\rho)\)].
In other words, a barotrope has no buoyancy, because the squared Brunt-V\"ais\"al\"a frequency \(N^2 \propto \bm{A}\cdot \nabla P_0\) vanishes.
The barotropic approximation is valid, when the Coriolis force dominates buoyancy, with \(N \lesssim |\bm{\Omega}|\).

The Brunt-V\"ais\"al\"a frequency can be estimated by \(N \sim (x/2)^{1/2} g c_{\text{s}}^{-1}\), where \(x = 6 \times 10^{-3} \rho / \rho_{\text{nuc}}\) is the local ratio of protons to neutrons in chemical equilibrium, \(\rho_{\text{nuc}} = 2.7 \times 10^{14} \, \text{g cm}^{-3}\) is the nuclear saturation density, \(g\) is the local gravitational acceleration, and \(c_{\text{s}}\) is the adiabatic sound speed \citep{ReiseneggerGoldreich1992,LockitchEtAl2003}.
In a recent study, \citet{AltiparmakEtAl2022} reported \(c_{\text{s}}^2 / c^2 > 1/3\) at \(\rho \gtrsim 2\rho_{\text{nuc}}\) for most of the EOSs they consider, which parametrise \(c_{\text{s}}\) as a function of the chemical potential in the range \(1.1 \rho_{\text{nuc}} < \rho \lesssim 40 \rho_{\text{nuc}}\).
This implies \(150\,\text{Hz} \lesssim N \, (\rho / \rho_{\text{nuc}})^{-1/2} (g / 10^{14} \text{cm s}^{-2})^{-1} \lesssim 260\)\,Hz.
Therefore, the barotropic approximation \(N \lesssim |\bm{\Omega}|\) is valid at the margin for accreting millisecond X-ray pulsars with \(|\bm{\Omega}| / 2\pi \gtrsim 200\)\,Hz \citep{PatrunoWatts2021}.
This is consistent with the findings of \citet{PassamontiEtAl2009}, who demonstrated numerically that the difference between \(l=|m|\) $r$-mode frequencies in barotropic and non-barotropic stars is negligible.
In this paper, we assume that the barotropic condition is fulfilled for $r$-modes for simplicity.
A more realistic study of the general problem in nonbarotropic stars is left for future work.

\subsection{Slow-rotation expansion of eigenmodes}
\label{subsec2.3:Slow_rotation_expansion}
The normal modes of oscillation \(\bm{\xi}_{\alpha}(\bm{x}, t)\) of equation~\eqref{eqn:Euler equation perturbed}, summing to give the total displacement vector \(\bm{\xi} = \sum_{\alpha} \bm{\xi}_{\alpha}\), exhibit a harmonic time dependence, 
\begin{align}
    \bm{\xi}_{\alpha}(\bm{x}, t) = \bm{\xi}_{\alpha}(\bm{x}) e^{i \sigma_{\alpha} t} ,
\end{align}
because operators \(\mathbfss{B}\) and \(\mathbfss{C}\) are time-independent.
The subscript \(\alpha\) labels the eigenmodes.
The eigenvalue \(\sigma_{\alpha} = \omega_{\alpha} + i/\tau_{\alpha}\) is complex in general, and its imaginary part \(1/\tau_{\alpha}\) describes the damping or growth rate of the mode.
As the system described by equations~\eqref{eqn:mass continuity rot_frame}--\eqref{eqn:Poisson equation rot_frame} involves no dissipation (e.g.\ ideal fluid, adiabatic oscillations), the eigenvalues \(\sigma_{\alpha} = \omega_{\alpha}\) are real.

We adopt the slow-rotation approximation and expand the eigenfrequencies and mode functions in orders of \(\Omega = |\bm{\Omega}|\) in the corotating frame \citep{SchenkEtAl2001}, viz.
\begin{align}
    \omega_{\alpha} = \omega_{\alpha}^{(0)} + \omega_{\alpha}^{(1)} + \omega_{\alpha}^{(2)} + \Order{\Omega^3} ,
    \label{eqn:eigenfrequency_expansion_in_Omega}
    \\
    \bm{\xi}_{\alpha} = \bm{\xi}_{\alpha}^{(0)} + \bm{\xi}_{\alpha}^{(1)} + \bm{\xi}_{\alpha}^{(2)} + \Order{\Omega^3} ,
    \label{eqn:mode_function_expansion_in_Omega}
\end{align}
where the bracketed superscript denotes the order in \(\Omega\). The zeroth-order terms correspond to eigenfrequencies and mode functions in the non-rotating limit. 
The operator \(\mathbfss{B}\) by definition [equation~\eqref{eqn:operator B}] is of order \(\Omega\), i.e., \(\mathbfss{B} = \mathbfss{B}^{(1)}\).
The operator \(\mathbfss{C}\) is expanded as \(\mathbfss{C} = \mathbfss{C}^{(0)} + \mathbfss{C}^{(2)} + \Order{\Omega^4}\), as implied by equation~\eqref{eqn:Euler equation background}.

The frequencies of $r$-modes vanish in the non-rotating limit, i.e.\ \(\omega_{\alpha}^{(0)} =  \mathbfss{C}^{(0)} \cdot \bm{\xi}^{(0)}_{\alpha} = 0\). The non-trivial equation at order \(\Omega^2\) reads
\begin{align} 
    0 = \left[- \omega_{\alpha}^{(1)2} + i \omega_{\alpha}^{(1)} \mathbfss{B}^{(1)} + \bm{C}^{(2)}\right] \cdot \bm{\xi}_{\alpha}^{(0)} + \bm{C}^{(0)} \cdot \bm{\xi}_{\alpha}^{(2)} .
    \label{eqn:second_order_eigenvalue_problem}
\end{align}
The component of equation~\eqref{eqn:second_order_eigenvalue_problem} projected into the subspace spanned by zero-frequency modes \(\bm{\xi}_{\alpha}^{(0)}\) in a non-rotating barotrope gives \citep{SchenkEtAl2001}
\begin{align}
    0 = \nabla \times \left[
        - \omega_{\alpha}^{(1)2}\bm{\xi}_{\alpha}^{(0)} + i \omega_{\alpha}^{(1)} \mathbfss{B}^{(1)} \cdot \bm{\xi}_{\alpha}^{(0)}
    \right] ,
    \label{eqn:second_order_eigenvalue_problem_2}
\end{align}
which is sufficient to determine \(\omega_{\alpha}^{(1)}\) and \(\bm{\xi}_{\alpha}^{(0)}\).
Equation~\eqref{eqn:second_order_eigenvalue_problem_2} can also be derived equivalently from vorticity conservation \citep{FriedmanStergioulas2013}.

In slowly-rotating barotropic stars, $r$-modes exist only for \(l=|m|\) in Newtonian gravity \citep{FriedmanStergioulas2013}, where \((l,m)\) are the orders of the spherical harmonic \(Y_{lm}\), and we write \(\alpha = (l, m)\).
For purely axial $r$-modes, one obtains
\begin{align}
    \bm{\xi}_{\alpha}^{(0)} = U_{lm}(r) (\hat{\bm{r}} \times \nabla)Y_{lm} ,
    \label{eqn:mode_function_r_mode_general}
\end{align}
where \(U_{lm}\) is a radial eigenfunction, and \(\hat{\bm{r}}\) is the radial unit vector. 
The rotational distortion of the stellar surface into an ellipse affects \(U_{lm}\) at order \(\Omega^2\).
Upon substituting equation~\eqref{eqn:mode_function_r_mode_general} into equation~\eqref{eqn:second_order_eigenvalue_problem_2}, we arrive at
\begin{align}
    \omega_{\alpha}^{(1)} = \frac{2 m \Omega}{l(l+1)} ,
    \label{eqn:omega_r_mode}
\end{align}
and
\begin{align}
    U_{lm}(r) \propto r^{l+1} ,
\end{align}
for \(l = |m|\), and \(U_{lm} = 0\) otherwise. 
The derivation is detailed in Appendix~\ref{appA:derviation_r_mode}.
Following the convention in \citet{Owen2010}, we normalise the Eulerian velocity perturbation of $r$-modes, \(\delta \bm{u}_{\!\alpha}(\bm{x}) = i \omega_{\alpha} \bm{\xi}_{\alpha}(\bm{x})\), according to
\begin{align}
    \delta \bm{u}_{\!\alpha}(\bm{x}) = \Omega R_* \left(\frac{r}{R}\right)^{|m|} (\bm{r} \times \nabla)Y_{|m|m} ,
    \label{eqn:r-mode_velocity_perturbation}
\end{align}
where \(R_*\) is the radius of the star.

\subsection{Orthogonality}
\label{subsec2.4:Orthogonality}
Eigenfunctions in rotating stars are not orthogonal in the conventional sense familiar from non-rotating stars. 
We define a symplectic product \(W\) of two vector functions \citep{FriedmanSchutz1978a}
\begin{align}
    W(\bm{\xi}_{\alpha}, \bm{\xi}_{\beta})
    = \left\langle \bm{\xi}_{\alpha}, \frac{\partial \bm{\xi}_{\beta}}{\partial t} + \frac{1}{2} \mathbfss{B} \cdot \bm{\xi}_{\beta} \right\rangle - \left\langle \frac{\partial \bm{\xi}_{\alpha}}{\partial t} + \frac{1}{2} \mathbfss{B} \cdot \bm{\xi}_{\alpha},\, \bm{\xi}_{\beta} \right\rangle ,
    \label{eqn:symplectic product}
\end{align}
where the angular brackets denote the inner product defined by
\begin{align}
    \langle \bm{\xi}_{\alpha}, \bm{\xi}_{\beta} \rangle 
    = \int \text{d}^3 x \, \rho_0 \bm{\xi}_{\alpha}^{*} \cdot \bm{\xi}_{\beta} .
    \label{eqn:inner product}
\end{align}
The symplectic product \(W(\bm{\xi}_{\alpha}, \bm{\xi}_{\beta})\) is time-independent (\(\text{d}W/\text{d}t = 0\)) for \(\bm{\xi}_{\alpha}\) and \(\bm{\xi}_{\beta}\) satisfying equation~\eqref{eqn:Euler equation perturbed}. 
When one has \(\omega_{\alpha} \neq \omega_{\beta}\) for \({\alpha}\neq {\beta}\), \(\text{d}W(\bm{\xi}_{\alpha}, \bm{\xi}_{\beta})/\text{d}t=0\) implies the modified orthogonality condition \citep{SchenkEtAl2001,PnigourasEtAl2024}
\begin{align}
    (\omega_{\alpha} + \omega_{\beta}) \langle \bm{\xi}_{\alpha}, \bm{\xi}_{\beta} \rangle - \langle \bm{\xi}_{\alpha}, i \mathbfss{B} \cdot \bm{\xi}_{\beta} \rangle = \mathcal{N}_{\alpha} \delta_{\!\alpha \beta} ,
    \label{eqn:modified orthogonality condition}
\end{align}
where \(\delta_{\!\alpha \beta}\) is the Kronecker delta, \(\mathcal{N}_{\alpha}\) is a normalisation constant, and the Einstein summation convention is suppressed temporarily on the right-hand side of equation~\eqref{eqn:modified orthogonality condition}. 
Similarly, we have \(W(\bm{\xi}_{\alpha}^{*}, \bm{\xi}_{\alpha}) = 0\).
Following \citet{FriedmanEtAl2017}, we write \(\mathcal{N}_{\alpha}\) as
\begin{align}
    \mathcal{N}_{\alpha} = 2 \omega_{\alpha} \kappa_{\alpha} \langle \bm{\xi}_{\alpha}, \bm{\xi}_{\alpha} \rangle ,
\end{align}
with
\begin{align}
    \kappa_{\alpha} = 1 - \frac{\Omega}{\omega_{\alpha}} \frac{\langle \bm{\xi}_{\alpha}, i \bm{\hat{\Omega}} \times \bm{\xi}_{\alpha} \rangle}{\langle \bm{\xi}_{\alpha}, \bm{\xi}_{\alpha} \rangle} .
\end{align}
For $r$-modes, the leading slow-rotation order of \(\kappa_{\alpha}\) is \(\kappa_{\alpha}^{(0)} = 1/2\) (see Appendix~\ref{appA:derviation_r_mode}), and the leading order of \(\mathcal{N}_{\alpha}\) is
\begin{align}
    \mathcal{N}_{\alpha}^{(1)} 
    &= \omega_{\alpha}^{(1)} \langle \bm{\xi}_{\alpha}^{(0)}, \bm{\xi}_{\alpha}^{(0)} \rangle^{(0)} ,
    \label{eqn:normalisation_constant^(1)}
\end{align}
where \(\langle \cdot \,,\, \cdot \rangle^{(0)}\) denotes the inner product in the non-rotating limit, as introduced in Appendix~\ref{appA:derviation_r_mode}, and the Einstein summation convention is suppressed temporarily again in equation~\eqref{eqn:normalisation_constant^(1)}.

\section{Excitation by accretion}
\label{sec:3-Accretion_excitation}
In this section, we calculate the response of $r$-modes to the mechanical impact of accreted matter.
We assume that the accretion occurs stochastically as a random sequence of discrete clumps of matter, as implied by modern numerical simulations \citep{RomanovaOwocki2015}.
In Section~\ref{subsec3.1:General_solution_inhomogeneous_problem}, we formulate the associated inhomogeneous boundary value problem and solve it in terms of a Green's function.
The force densities corresponding to a single clump and a stochastic sequence of clumps with random arrival times are defined in Sections~\ref{subsec3.2:Force_density_single_clump} and \ref{subsec3.3:Stochastic_sequence_of_clumps}, respectively.
We also calculate the temporal autocorrelation function of the stochastic response in Section~\ref{subsec3.4:Temporal_autocorrelation_function}, to lay the foundation for calculating the root-mean-square wave strain and power spectral density of the emitted gravitational radiation in Section~\ref{sec:4-gravitational_radiation}.

\subsection{General solution to the inhomogeneous problem}
\label{subsec3.1:General_solution_inhomogeneous_problem}
The inhomogeneous excitation problem can be formulated as
\begin{align}
    \frac{\partial^2 \bm{\xi}(\bm{x}, t)}{\partial t^2} + \mathbfss{B} \cdot \frac{\partial \bm{\xi}(\bm{x}, t)}{\partial t} + \mathbfss{C} \cdot \bm{\xi}(\bm{x}, t)
    = \frac{\bm{\mathcal{F}}(\bm{x}, t)}{\rho_0} ,
    \label{eqn:Inhomogeneous excitation problem}
\end{align}
where \(\bm{\mathcal{F}}(\bm{x}, t)\) is the mechanical force density from accretion impacts.
The standard mode decomposition of \(\bm{\xi}(\bm{x}, t)\) \citep{SchenkEtAl2001,PnigourasEtAl2024,DongMelatos2024} couples together the coefficients in equation~\eqref{eqn:Inhomogeneous excitation problem} for distinct modes \(\alpha\) and \(\alpha' \neq \alpha\) in a rotating star.
Therefore, it is convenient to recast equation~\eqref{eqn:Inhomogeneous excitation problem} as \citep{DysonSchutz1979}
\begin{align}
    \frac{\partial}{\partial t} \begin{bmatrix}
        \bm{\xi}(\bm{x}, t) \\
        \bm{\dot{\xi}}(\bm{x}, t)
    \end{bmatrix}
    = \begin{bmatrix}
        0 & 1 \\
        -\mathbfss{C} & -\mathbfss{B}
    \end{bmatrix} \begin{bmatrix}
        \bm{\xi}(\bm{x}, t) \\
        \bm{\dot{\xi}}(\bm{x}, t)
    \end{bmatrix} + \begin{bmatrix}
        0 \\
        \bm{\mathcal{F}}(\bm{x}, t)/\rho_0
    \end{bmatrix} .
    \label{eqn:phase_space_excitation_problem}
\end{align}
Hence the orthogonality condition \eqref{eqn:modified orthogonality condition} for the column matrix \(\bm{\chi}_{\alpha} = [\bm{\xi}_{\alpha}, \bm{\dot{\xi}}_{\alpha}]^{\text{T}}\) becomes
\begin{align}
    \int \text{d}^3x \, i \rho_0 \bm{\chi}_{\alpha}^{\dagger} \mathbfss{W} \bm{\chi}_{\beta}
    = \mathcal{N}_{\alpha} \delta_{\!\alpha \beta} ,
    \label{eqn:orthogonality_condition_phase_space}
\end{align}
where the superscript \(^{\text{T}}\) denotes the transpose, the dagger denotes the conjugate transpose, and the symplectic matrix \(\mathbfss{W}\) is defined by \citep{DysonSchutz1979}
\begin{align}
    \mathbfss{W} = \begin{bmatrix}
        -\mathbfss{B} & -1 \\
        1 & 0
    \end{bmatrix} .
\end{align}
We expand \(\bm{\chi}\) according to the following ansatz \citep{SchenkEtAl2001,PnigourasEtAl2024},
\begin{align}
    \begin{bmatrix}
        \bm{\xi}(\bm{x}, t) \\
        \bm{\dot{\xi}}(\bm{x}, t)
    \end{bmatrix}
    = \sum_{\alpha} c_{\alpha}(t) \begin{bmatrix}
        \bm{\xi}_{\alpha}(\bm{x}) \\
        \delta \bm{u}_{\!\alpha}(\bm{x})
    \end{bmatrix} + \text{c.c.} ,
    \label{eqn:phase_space_expansion}
\end{align}
where c.c.\ denotes the complex conjugate. Upon substituting equation~\eqref{eqn:phase_space_expansion} into equation~\eqref{eqn:phase_space_excitation_problem} and applying equation~\eqref{eqn:orthogonality_condition_phase_space}, we arrive at the following equations of motion for \(c_{\alpha}(t)\) and \(c_{\alpha}^{*}(t)\), 
\begin{align}
    \frac{\text{d}\,c_{\alpha}(t)}{\text{d}t} - i \omega_{\alpha} c_{\alpha}(t) 
    &= -\frac{i \langle \bm{\xi}_{\alpha}(\bm{x}), \bm{\mathcal{F}}(\bm{x},t)/\rho_0 \rangle}{\mathcal{N}_{\alpha}} ,
    \label{eqn:c_alpha_DE}
\end{align}
Equation~\eqref{eqn:c_alpha_DE} is diagonalised, in the sense that there is no coupling between the equations governing distinct modes \(\alpha\) and \(\alpha' \neq \alpha\).
The solution to equation~\eqref{eqn:c_alpha_DE} is given by
\begin{align}
    c_{\alpha}(t) = -\frac{i}{\mathcal{N}_{\alpha}} 
    \int \text{d}^3x \int_{t_0}^t \text{d}t' \, \bm{\xi}_{\alpha}^{*}(\bm{x}) \cdot \bm{\mathcal{F}}(\bm{x},t') e^{i \omega_{\alpha} (t-t')} ,
    \label{eqn:c_alpha_general_solution}
\end{align}
assuming that the initial condition is \(c_{\alpha}(t) = 0\) for \(t<t_0\).
We set \(t_0 = 0\) without loss of generality.
The solution \(c_{\alpha}^{*}(t)\) is obtained by taking the complex conjugate of equation~\eqref{eqn:c_alpha_general_solution}.
Equation~\eqref{eqn:c_alpha_general_solution} gives an explicit expression for the mode amplitude \(c_{\alpha}(t)\) excited by the force density \(\bm{\mathcal{F}}(\bm{x}, t)\) and is a key input into the GW calculations in Section~\ref{sec:4-gravitational_radiation}.

Real neutron stars have viscosity. We follow standard practice \citep{OwenEtAl1998,KokkotasEtAl2001,HoEtAl2020} and incorporate phenomenological damping or growth factors \(\propto \exp[-(t-t')/\tau_{\alpha}]\) in equation~\eqref{eqn:c_alpha_general_solution} to account for viscous damping or growth due to the back reaction from gravitational radiation; see Section~\ref{sec:4-gravitational_radiation}.

\subsection{Force density from a single clump}
\label{subsec3.2:Force_density_single_clump}
We denote the force density from the impact of a single clump starting at \(t_{\rm s}\) as \(\bm{\mathcal{F}}(\bm{x}, t; t_{\rm s})\).
For simplicity, we model \(\bm{\mathcal{F}}(\bm{x},t; t_{\rm s})\) to be a Dirac delta function in position (i.e.\ the idealised clump strikes the surface at a single point) and constant in time during its interaction with the stellar surface; see Section~4 in \citet{DongMelatos2024} and references therein for a physical justification. We then write
\begin{align}
    \bm{\mathcal{F}}(\bm{x}, t; t_{\rm s}) = \bm{p}_{\rm s} T_{\rm s}^{-1} \delta(\bm{x}-\bm{x}_{\rm s}) [H(t - t_{\rm s}) - H(t - t_{\rm s} - T_{\rm s})] ,
    \label{eqn:force_density_single_pulse}
\end{align}
where \(\bm{p}_{\rm s}\) is the momentum of the clump in the corotating frame, \(T_{\rm s}\) is the duration of the impact, \(\bm{x}_{\rm s}\) is the impact point on the surface of the star, and \(H(\cdot)\) denotes the Heaviside step function.

The force density model in equation~\eqref{eqn:force_density_single_pulse} is highly idealised.
For example, non-polar-cap accretion flows and inhomogeneous hotspot geometries are expected to produce more complicated spatial profiles of the force density \citep{RomanovaEtAl2004,KulkarniRomanova2008,RomanovaOwocki2015}.
Improvements to the model could include extended radial and angular profiles corresponding to the aforementioned geometries, as well as oscillatory temporal profiles.
The extended radial and angular profiles are likely to reduce the $r$-mode amplitude due to spatial averaging.
The reduction is predicted to be smaller for $r$-modes, whose eigenfunctions are nodeless and monotonic in \(r\), than for high-radial-order $p$-modes, whose eigenfunctions oscillate in \(r\).
Temporal oscillations affect the $r$-mode amplitude negligibly, as long as \(T_{\rm s} \lesssim \omega_{\alpha}^{-1}\) is satisfied.
This is predicted to be the case even near resonance, where the oscillation frequency of the force density is close to \(\omega_{\alpha}\), because the impact duration is too short for the oscillatory force density to go through many full cycles.
Equation~\eqref{eqn:force_density_single_pulse} with \(T_{\rm s} \sim 10^{-4}\,\)s replicates the $f$-mode dominance in the GW energy spectrum emitted by an infalling, quadrupolar mass shell simulated by \citet{NagarEtAl2004} \citep[see Section~5.3 in][]{DongMelatos2024}.
Nevertheless, it is straightforward to generalise the analysis to a more distributed (and hence more realistic) \(\bm{\mathcal{F}}(\bm{x},t; t_{\rm s})\) through equation~\eqref{eqn:c_alpha_general_solution}, which expresses the mode response in terms of the Green function response to an impulsive excitation convolved with \(\bm{\mathcal{F}}(\bm{x},t; t_{\rm s})\).

\subsection{Stochastic sequence of clumps}
\label{subsec3.3:Stochastic_sequence_of_clumps}
In real astrophysical systems, accretion is stochastic.
The aperiodic X-ray flux is observed to fluctuate at frequencies \(\sim \text{kHz}\) in accreting neutron stars \citep{SunyaevRevnivtsev2000}, largely due to mass accretion rate fluctuations arising from clumps in the flow.
Clumps arise naturally in inhomogeneous accretion flows through self-healing magnetic and radiation-driven Rayleigh-Taylor instabilities \citep{MorfillEtAl1984,Wang1987a,DemmelEtAl1990,SpruitTaam1993,DAngeloSpruit2010}.
Quasi-periodic oscillations in X-ray lightcurves may be associated with beat frequencies between the angular velocities of the neutron star and the clumps \citep{LambEtAl1985}.
Correlations between data in the near-IR and optical bands provide evidence of accreted plasma clumps in PSR J1023+0038 \citep{ShahbazEtAl2018}.
Typical X-ray pulsars have root-mean-square fractional variabilities of \(\sim 20\%\) in their X-ray fluxes \citep{BelloniHasinger1990}.
The observed power spectral density of the X-ray flux in many objects is white below the turnover frequency and red above the turnover frequency, with power-law indices \(\approx 1\)--\(1.5\) \citep{HoshinoTakeshima1993,LazzatiStella1997}. 
The turnover frequency approaches \(\Omega\) for systems with magnetospheric radius close to the corotation radius \citep{RevnivtsevEtAl2009}.

State-of-the-art, three-dimensional, magnetohydrodynamic simulations show that the accretion rate fluctuates on a time-scale of milliseconds in response to self-healing Rayleigh-Taylor and Kelvin-Helmholtz instabilities at the disk-magnetosphere boundary \citep{RomanovaEtAl2008,RomanovaOwocki2015,MushtukovEtAl2024}.
The millisecond variability of the accretion rate is consistent with the expected timescale of the fragmentation of the clumpy accretion stream at the Keplerian period \(\Omega_{\rm K}^{-1}\) near the disk-magnetosphere boundary \(r = r_{\rm m}\), with \(\Omega_{\rm K}(r_{\rm m})^{-1} = (GM / r_{\rm m}^3)^{-1/2} = 2.7 (M / \Msun)^{-1/2} (r_{\rm m} / 100\,\text{km})^{3/2}\)\,ms.
It is also broadly consistent with the propagating fluctuation model for a geometrically thick accretion disk (\(H / r_{\rm m} \sim 1\)), in which clumps propagate inwards on the viscous timescale in the inner emission region \citep{Lyubarskii1997}. The viscous timescale at \(r = r_{\rm m}\) is estimated as \(t_{\rm visc}(r_{\rm m}) \propto \alpha_{\rm disk}^{-1} (H / r_{\rm m})^{-2} \Omega_{\rm K}(r_{\rm m})^{-1}\), where \(\alpha_{\rm disk} \leq 1\) is the viscosity parameter, and \(H / r_{\rm m}\) is the disk scale height to radius ratio \citep{Lyubarskii1997}.
The stochasticity enters equation~\eqref{eqn:force_density_single_pulse} through random arrival times \(t_{\rm s}\), random impact points \(\bm{x}_{\rm s}\), the random impact duration \(T_{\rm s}\), and the random clump momentum \(\bm{p}_{\rm s}\) \citep{DongMelatos2024}.
However, it is difficult to resolve individual clumps observationally.
In this paper, for the sake of simplicity, we assume that every clump has the same momentum and impact location, and that every impact has the same duration.
We model the stochasticity in \(t_{\rm s}\) according to
\begin{align}
    \bm{\mathcal{F}}(\bm{x}, t) = \sum_{\text{s}=1}^{N(t)} \bm{\mathcal{F}}(\bm{x}, t; t_{\rm s}) ,
    \label{eqn:force_density_stochastic_pulse}
\end{align}
where \(\{t_{\rm s}\}\) is a sequence of impact epochs drawn from a homogeneous Poisson point process with mean clump impact rate \(f_{\rm acc}\), and \(N(t)\) counts the total (and random) number of impacts up to time \(t\). 

Equation~\eqref{eqn:force_density_stochastic_pulse} is consistent with the standard shot noise model \citep{Terrell1972,LambEtAl1985,LochnerEtAl1991}, with shot noise profiles in accretion disks replaced by \(\bm{\mathcal{F}}(\bm{x}, t; t_{\rm s})\).
The shot noise model reproduces the observed red-noise power spectrum in some (although not all) sources \citep{LazzatiStella1997} and is representative enough to be used as a first pass to study the stochastic excitation of oscillations by accretion.
It is also worth noting that the shot noise model struggles to reproduce the observed proportionality between rms variability and X-ray flux on short time-scales in some sources \citep{UttleyMcHardy2001}.
Generalising the analysis to more sophisticated models, such as the propagating fluctuation model \citep{Lyubarskii1997}, is left for future work.

\subsection{Temporal autocorrelation function}
\label{subsec3.4:Temporal_autocorrelation_function}
The autocorrelation function of \(c_{\alpha}(t)\) excited by the force density in equation~\eqref{eqn:force_density_stochastic_pulse} can be calculated from
\begin{align}
    \left\langle c_{\alpha}(t) c_{\beta}^{*}(t') \right\rangle _{\rm s}
    =& f_{\rm acc}^2 \int_{0}^{t} \text{d}t_{\rm s} \int_{0}^{t'} \text{d}t_{\rm s}' \, 
    \langle c_{\alpha}(t; t_{\rm s}) \rangle _{\rm s}
    \langle c_{\beta}^{*}(t'; t_{\rm s}') \rangle _{\rm s}
    \nonumber \\
    &+ f_{\rm acc} \int_{0}^{\min\{t, t'\}} \text{d}t_{\rm s} \, 
    \left\langle c_{\alpha}(t; t_{\rm s}) c_{\beta}^{*}(t'; t_{\rm s}) \right\rangle _{\rm s} ,
    \label{eqn:autocorrelation_function_c_alpha_general}
\end{align}
where angular brackets with the subscript `s' denote the ensemble average over all random realisations of the stochastic impact process.
Equation~\eqref{eqn:autocorrelation_function_c_alpha_general} is derived in Appendix~D in \citet{DongMelatos2024}.
For \(T_{\rm s}^{-1} \gg \omega_{\alpha}, \omega_{\beta}\), the first term on the right-hand side of equation~\eqref{eqn:autocorrelation_function_c_alpha_general} reduces to
\begin{align}
    &f_{\rm acc}^2 
    \int_{0}^{t} \text{d}t_{\rm s} \int_{0}^{t'} \text{d}t_{\rm s}' \, 
    \langle c_{\alpha}(t; t_{\rm s}) \rangle_{\rm s}
    \langle c_{\beta}^{*}(t'; t_{\rm s}') \rangle_{\rm s}
    \nonumber \\
    &\propto 
    f_{\rm acc}^2 \tau_{\alpha} \tau_{\beta}
    [(\omega_{\alpha} \tau_{\alpha} + i) \mathcal{N}_{\alpha}]^{-1} 
    [1 - \exp(-t/\tau_{\alpha}) \exp(i \omega_{\alpha} t)]
    \nonumber \\ &\quad \times
    [(\omega_{\beta} \tau_{\beta} - i) \mathcal{N}_{\beta}]^{-1} 
    [1 - \exp(-t'/\tau_{\beta})\exp(- i \omega_{\beta} t')] .
    \label{eqn:autocorrelation_facc_square}
\end{align}
Notably, \(T_{\rm s}\) does not appear in equation~\eqref{eqn:autocorrelation_facc_square}, as an impact with \(T_{\rm s}^{-1} \gg \omega_{\alpha}, \omega_{\beta}\) approximately behaves, as if the mode is excited by a delta-function impulse.
Equation~\eqref{eqn:autocorrelation_facc_square} tends to a constant for \(\tau_{\alpha} > 0\) and \(\tau_{\beta} > 0\), as \(t\) and \(t'\) go to infinity.
The sign of \(\tau_{\alpha}\) depends on whether or not the CFS instability is triggered; see Section~\ref{sec:5-CFS_instability}.
The second term on the right-hand side of equation~\eqref{eqn:autocorrelation_function_c_alpha_general} is negligible, unless the two modes have the same frequency.
With \(T_{\rm s}^{-1} \gg \omega_{\alpha}\) and \(\alpha = \beta\), the second term on the right-hand side of equation~\eqref{eqn:autocorrelation_function_c_alpha_general} reduces to
\begin{align}
    &f_{\rm acc} \int_{0}^{\min\{t, t'\}} \text{d}t_{\rm s} \, 
    \left\langle c_{\alpha}(t; t_{\rm s}) c_{\alpha}^{*}(t'; t_{\rm s}) \right\rangle _{\rm s}
    \nonumber \\
    &\propto 
    f_{\rm acc} \tau_{\alpha} \mathcal{N}_{\alpha}^{-2}
    \exp(-\zeta / \tau_{\alpha}) \exp(-i \omega_{\alpha} \zeta) 
    \nonumber \\ &\quad \times
    \left\{1 - \exp[-2\min(t, t') / \tau_{\alpha}]\right\} /2 \, ,
    \label{eqn:autocorrelation_facc_c_alpha_c_alpha_cc}
\end{align}
where we denote the time lag between \(t\) and \(t'\) by \(\zeta = |t - t'|\).
Equation~\eqref{eqn:autocorrelation_facc_c_alpha_c_alpha_cc} is stationary, i.e. it depends only on \(\zeta\), in the limit \(\min(t, t')\rightarrow \infty\).
The root-mean-square mode amplitude \(c_{\alpha,\text{rms}}\) can be evaluated by setting \(\alpha=\beta\) and \(\zeta = 0\) in equation~\eqref{eqn:autocorrelation_function_c_alpha_general}.

We note for completeness that the following result is also useful for calculating the root-mean-square GW strain in Section~\ref{sec:4-gravitational_radiation}:
\begin{align}
    &f_{\rm acc} \int_{0}^{\min\{t, t'\}} \text{d}t_{\rm s} \, 
    \left\langle c_{\alpha}(t; t_{\rm s}) c_{\alpha}(t'; t_{\rm s}) \right\rangle _{\rm s}
    \nonumber \\
    &\propto 
    - i f_{\rm acc} \tau_{\alpha} 
    \mathcal{N}_{\alpha}^{-2} [2 (i + \omega_{\alpha} \tau_{\alpha})]^{-1}
    \exp(-\zeta / \tau_{\alpha}) \exp(i \omega_{\alpha} \zeta) 
    \nonumber \\ &\quad \times
    \left\{1 - \exp[- 2\min(t, t') / \tau_{\alpha}] \exp[2i \omega_{\alpha} \min(t, t')]\right\} .
    \label{eqn:autocorrelation_facc_c_alpha_c_alpha}
\end{align}
Equation~\eqref{eqn:autocorrelation_facc_c_alpha_c_alpha} is related closely to equation~\eqref{eqn:autocorrelation_facc_c_alpha_c_alpha_cc}, except that we do not take the complex conjugate of \(c_{\alpha}(t'; t_{\rm s})\).

\section{Gravitational radiation}
\label{sec:4-gravitational_radiation}
In this section, we explore the stochastic GW signal emitted by the accretion-excited $r$-modes calculated in Section~\ref{sec:3-Accretion_excitation}.
We calculate the wave strain generated by modes excited by a single clump in Section~\ref{subsec4.1:Strain}. 
In Sections~\ref{subsec4.2:Root_mean_square_wave_strain_without_CFS_instabilities} and \ref{subsec4.3:Power_spectral_density}, we calculate the root-mean-square wave strain and the power spectral density of the emitted gravitational radiation, respectively, in order to investigate its detectability.

\subsection{Instantaneuous wave strain}
\label{subsec4.1:Strain}
The metric perturbation \(h_{jk}^{\rm TT}\) in the transverse traceless gauge from Newtonian $r$-modes, measured by an observer at a distance \(d\), is given by a linear combination of the time-varying current multipole moments \citep{Thorne1980},
\begin{align}
    h_{jk}^{\rm TT}(t) = 
        \sum_{l=2}^{\infty}
        \frac{G}{c^{3+l} d} 
        \sum_{m=-l}^{l}
        \frac{\text{d}^{l} \delta S_{lm}(t)}{\text{d} t^{l}}
        T^{\text{B}2,lm}_{jk} ,
    \label{eqn:metric_perturbation}
\end{align}
with \citep{MelatosPeralta2009}
\begin{align}
    \delta S_{lm} (\delta \bm{u})
    =&
    - \frac{32\pi}{(2l+1)!!} \left[\frac{l+2}{2l(l-1)(l+1)}\right]^{1/2}
    \nonumber \\ &\times
    \delta \int \text{d}^3 x \, r^{l} \bm{x} \cdot \text{curl}(\rho \bm{v}) Y_{lm}^{*}
    \label{eqn:S_lm_form2}
    \\
    =&
    - \frac{32\pi}{(2l+1)!!} \left[\frac{l+2}{2l(l-1)(l+1)}\right]^{1/2}
    \nonumber \\ &\times
    \int \text{d}^3 x \, r^{l} \bm{x} \cdot \text{curl}(\rho \delta \bm{u}) Y_{lm}^{*} ,
    \label{eqn:deltaS_lm_form2}
\end{align}
where \(t\) is the retarded time, and \(T^{\text{B}2,lm}_{jk}\) is the gravitomagnetic tensor describing the angular beam pattern associated with the \((l, m)\) spherical harmonic.
Equation~\eqref{eqn:deltaS_lm_form2} follows from equation~\eqref{eqn:S_lm_form2}, because one has \(\delta \bm{u} = \delta \bm{v}\) and \(\delta \rho = 0\) for purely axial $r$-modes.

We write equation~\eqref{eqn:metric_perturbation} in terms of \(h_0(t; t_{\rm s}, l)\), the strain generated in the \(l\)-th multipole moment by a mode \(c_{\alpha}(t; t_{\rm s})\) excited by a single clump striking at time \(t_{\rm s}\), as 
\begin{align}
    h_{jk}^{\rm TT}(t) = 
    \sum_{l=2}^{\infty}
    T^{\text{B}2,l l}_{jk} 
    \sum_{\text{s}=1}^{N(t)}
    h_0(t; t_{\rm s}, l) 
    + \text{c.c.} \, ,
\end{align}
with
\begin{align}
    h_0(t; t_{\rm s}, l) =& 
    \frac{G}{c^{3+l} d} 
    \sum_{m=-l}^{l}
    \delta S_{lm}(\delta \bm{u}_{\!\alpha})
    \frac{\text{d}^{l} c_{\alpha}(t; t_{\rm s})}{\text{d} t^{l}} .
    \label{eqn:strain_single_impact_general}
\end{align}
Interestingly, \(h_0(t; t_{\rm s}, l)\) does not depend on the EOS, because \(\delta S_{lm}(\delta \bm{u}_{\!\alpha})\) cancels out \(1/\mathcal{N}_{\alpha}^{(1)}\) in \(c_{\alpha}(t; t_{\rm s})\), and \(\rho\) does not enter the integrand in equation~\eqref{eqn:c_alpha_general_solution}.
The cancellation occurs physically, because purely axial $r$-modes in a barotropic star have power-law radial profiles \(U_{lm}(r)\).
Evaluating equation~\eqref{eqn:strain_single_impact_general} assuming equation~\eqref{eqn:force_density_single_pulse}, as well as \(T_{\rm s}^{-1} \gg \omega_{\alpha}\) and \(\omega_{\alpha} \tau_{\alpha} \gg 1\), the strain \(h_0(t; t_{\rm s}, 2)\) from the dominant \(l=2\) multipole simplifies to 
\begin{align}
    h_0(t; t_{\rm s}, 2)
    =& \, 1.83 \times 10^{-40} \gamma_{v}
    \left(\frac{d}{1 \, \text{kpc}}\right)^{-1} 
    \left(\frac{R_{*}}{10 \, \text{km}}\right)^{2} 
    \left(\frac{\nu_{\rm s}}{10 \, \text{Hz}}\right)^{2} 
    \nonumber \\ &\times
    \left(\frac{\dot{M}}{10^{-8} \, \MsunPerYr}\right)
    \left(\frac{f_{\text{acc}}}{1 \, \text{kHz}}\right)^{-1}
    \left(\frac{|\bm{v}|}{0.4 c}\right)  
    \nonumber \\ &\times
    \left\{
    \hat{\bm{v}} \cdot \left[(\bm{x} \times \nabla)Y_{22}(\hat{\bm{x}})\right]_{\bm{x} = \bm{x}_{\rm s}} 
    \exp[i \omega_{\alpha, \text{i}} (t - t_{\rm s})] 
    + \text{c.c.} 
    \right\} 
    \nonumber \\ &\times
    \exp[-(t - t_{\rm s})/\tau_{\alpha}] ,
    \label{eqn:strain_single_impact}
\end{align}
where we write \(\gamma_{v} = [1 - (|\bm{v}|/c)^2]^{-1/2}\), \(\nu_{\rm s}\) is the spin frequency, \(\dot{M}\) is the accretion rate, \(\bm{v}\) is the clump velocity in the corotating frame, and \(\omega_{\alpha, \text{i}}\) represents the angular velocity in the inertial frame.
In deriving equation~\eqref{eqn:strain_single_impact}, we assume that the accretion rate is constant over the accretion episode and write \(\bm{p}_{\rm s} = \gamma_{v} m \bm{v}\) in the force density [equation~\eqref{eqn:force_density_single_pulse}], where \(m = \dot{M} / f_{\rm acc}\) is the typical mass of a clump.

The strain in equation~\eqref{eqn:strain_single_impact} scales \(\propto \nu_{\rm s}^{2}\). 
This differs from the \(\nu_{\rm s}^{3}\) scaling in \citet{Riles2023}, because the impact-excited $r$-mode amplitude scales \(\propto \nu_{\rm s}^{-1}\).
Physically, this occurs because the \(t'\) integral in equation~\eqref{eqn:c_alpha_general_solution} is independent of \(\nu_{\rm s}\) for \(T \ll \omega_{\alpha}^{-1}\), and we have \(c_{\alpha} \propto \mathcal{N}_{\alpha}^{-1} \propto \omega_{\alpha}^{-1} \propto \nu_{\rm s}^{-1}\), i.e.\ the mode energy \(E_{\alpha} \propto |c_{\alpha}|^{2} \nu_{\rm s}^{2}\) is independent of \(\nu_{\rm s}\).
The impact location \(\bm{x}_{\rm s}\) not only affects the amplitude, but also the phase through the \(e^{2i\phi}\) term in \(Y_{22}\). When \(\bm{x}_{\rm s}\) is fixed, the phase is set to zero without loss of generality.

\subsection{Root-mean-square wave strain}
\label{subsec4.2:Root_mean_square_wave_strain_without_CFS_instabilities}
In Sections~\ref{subsec4.2:Root_mean_square_wave_strain_without_CFS_instabilities} and \ref{subsec4.3:Power_spectral_density}, we assume that the viscous damping timescale \(\tau^{\text{visc}}_{\alpha}\) and gravitational radiation reaction timescale \(\tau^{\text{gw}}_{\alpha}\) are related by \(\tau^{\text{visc}}_{\alpha} < \left|\tau^{\text{gw}}_{\alpha}\right|\), so that the CFS instability is not triggered.
We define the temporal autocorrelation function of the strain as 
\(
    C_{\alpha}(t, t') = \left\langle h_{0}(t; \alpha) h_{0}^{*}(t'; \alpha) \right\rangle _{\rm s} ,
\)
with \(
    h_0(t; \alpha) = \sum_{t_\text{s} \leq t} h_0(t; t_{\rm s}, \alpha) .
\)
We calculate the root-mean-square characteristic wave strain, \(h_{\text{rms}}\), by evaluating \(C_{\alpha}(t, t')^{1/2}\) at zero lag (\(\zeta = 0\)), in order to assess the detectability of the stochastic signal.

In the regime \(\omega_{\alpha} \tau_{\alpha} \gg 1\), the dominant term of \(C_{\alpha}(t, t')\) reads
\begin{align}
    C_{\alpha}(t, t') =& \, 3.36 \times 10^{-80} \gamma_{v}^2
    \left(\frac{d}{1 \, \text{kpc}}\right)^{-2} 
    \left(\frac{R_{*}}{10 \, \text{km}}\right)^{4} 
    \left(\frac{\nu_{\rm s}}{10 \, \text{Hz}}\right)^{4} 
    \nonumber \\ &\times
    \left(\frac{\dot{M}}{10^{-8} \, \MsunPerYr}\right)^{2}
    \left(\frac{f_{\text{acc}}}{1 \, \text{kHz}}\right)^{-2}
    \left(\frac{|\bm{v}|}{0.4 c}\right)^{2}
    \nonumber \\ &\times
    \left| \hat{\bm{v}} \cdot \left[(\bm{x} \times \nabla)Y_{22}(\hat{\bm{x}})\right]_{\bm{x} = \bm{x}_{\rm s}} \right|^{2}
    K_{22}[\zeta; \min(t, t')] ,
    \label{eqn:K_h}
\end{align}
with
\begin{align}
    K_{\alpha}[\zeta; \min(t, t')] 
    =\,& f_{\rm acc} \tau_{\alpha} 
    \exp(-\zeta / \tau_{\alpha}) \cos(\omega_{\alpha} \zeta) 
    \nonumber \\ &\times
    \left\{1 - \exp[-2\min(t, t') / \tau_{\alpha}]\right\} .
    \label{eqn:C_2_autocorrelation}
\end{align}
Physically, \(K_{\alpha}(\zeta=0; t)\) equals the mean number of clumps striking the stellar surface up to time \(t\) for the mode \(\alpha\).
In the limit \(t \rightarrow \infty\), \(K_{\alpha}(\zeta=0; t)\) tends to \(f_{\rm acc} \tau_{\alpha}\), the saturated number of clumps during the damping timescale \(\tau_{\alpha}\).
In the regime \(\min(t, t') \ll \tau_{\alpha}\), \(K_{\alpha}(\zeta=0; t)\) reduces to \(2 f_{\rm acc} t\), and \(h_{\rm rms}\) depends on the duration of any particular uninterrupted accretion episode, with \(\Delta t_{\rm acc} \ll \tau_{\alpha}\).
Hence, we have
\begin{align}
    h_{\text{rms}} =& \, 4.61 \times 10^{-35} \gamma_{v}
    \left(\frac{d}{1 \, \text{kpc}}\right)^{-1} 
    \left(\frac{R_{*}}{10 \, \text{km}}\right)^{2} 
    \left(\frac{\nu_{\rm s}}{10 \, \text{Hz}}\right)^{2} 
    \nonumber \\ &\times
    \left(\frac{\dot{M}}{10^{-8} \, \MsunPerYr}\right)
    \left(\frac{f_{\text{acc}}}{1 \, \text{kHz}}\right)^{-1/2}
    \left(\frac{|\bm{v}|}{0.4 c}\right)
    \left(\frac{\Delta t_{\rm acc}}{1 \, \text{yr}}\right)^{1/2}
    \nonumber \\ &\times
    \left| \hat{\bm{v}} \cdot \left[(\bm{x} \times \nabla)Y_{22}(\hat{\bm{x}})\right]_{\bm{x} = \bm{x}_{\rm s}} \right| .
    \label{eqn:h_rms_unsaturated}
\end{align}
One can easily obtain the saturated \(h_{\rm rms}\), by replacing \(\Delta t_{\rm acc}\) with \(\tau_{\alpha} / 2\) in equation~\eqref{eqn:h_rms_unsaturated}.

The impact velocity direction \(\hat{\bm{v}}\) and the impact location \(\bm{x}_{\rm s}\) play important roles in determining \(h_{\rm rms}\).
For example, when taking \(\hat{\bm{v}} = \hat{\bm{\theta}}\) in equation~\eqref{eqn:h_rms_unsaturated}, we find \(\left| \hat{\bm{v}} \cdot \left[(\bm{x} \times \nabla)Y_{22}(\hat{\bm{x}})\right]_{\bm{x} = \bm{x}_{\rm s}}  \right|\) attains its maximum value of \(0.77\) at the equator.
As \(v_{\phi}\) increases, \(h_{\rm rms}\) decreases.
The impact location \(\bm{x}_{\rm s}\) that maximises \(h_{\rm rms}\) shifts towards the pole for \(\hat{v}_{\phi} > \hat{v}_{\theta}\). 
It maximises \(h_{\rm rms}\) at \(\theta = \pi/4\), with \(\left| \hat{\bm{v}} \cdot \left[(\bm{x} \times \nabla)Y_{22}(\hat{\bm{x}})\right]_{\bm{x} = \bm{x}_{\rm s}}  \right| = 0.39\), when \(\hat{\bm{v}} = \hat{\bm{\phi}}\). 

Equation~\eqref{eqn:h_rms_unsaturated} implies that GWs emitted by impact-excited $r$-mode oscillations in neutron stars are too weak to be detected by the current generation of LIGO detectors.
In an optimistic scenario, with \(v_{\rm s}=0.5\)kHz and \(\Delta t_{\rm acc}=10^{6}\)yr, one finds \(h_{\rm rms} \lesssim 1 \times 10^{-28}\), which corresponds to \(|c_{\alpha}| \lesssim 10^{-8}\) \citep{Riles2023}.
These values of \(h_{\rm rms}\) and \(|c_{\alpha}|\) are below the upper bounds \(h_{0}^{95\%} \sim 10^{-25}\) (at 95\% confidence level) inferred from recent searches of continuous waves from accreting systems \citep{MiddletonEtAl2020,CovasEtAl2022,LIGOScientificCollaborationEtAl2022} and the theoretically predicted amplitude set by nonlinear mode coupling, viz. \(10^{-7} \lesssim |c_{\alpha}| \lesssim 10^{-4}\) \citep{BondarescuEtAl2007,GusakovEtAl2014a}.

The impact-excited $f$- and $p$-modes in a non-rotating, barotropic star emit GWs with \(h_{\rm rms} \lesssim 10^{-33}\) \citep{DongMelatos2024}.
The $f$- and $p$-modes are strongly averaged during the top-hat impact.
On the other hand, the $g$-modes, like $r$-modes, are less affected by the top-hat averaging.
The GW strain emitted by $g$-modes varies from \(10^{-34}\) to \(10^{-32}\), depending on \(\Delta t_{\rm acc}\) and \(\hat{\bm{v}}\) \citep{DongMelatos2024}.
There are also differences between the EOS dependence of $f$- and $p$-modes and that of $r$-modes in a barotropic star. 
For instance, the GWs from $f$- and $p$-modes are weaker for a stiffer EOS (larger \(\partial \log{P} / \partial \log{\rho}\)) \citep{DongMelatos2024}. 
In contrast, the GWs from impact-excited $r$-modes are approximately independent of the EOS, as explained in Section~\ref{subsec4.1:Strain}, except that the EOS enters \(h_{\rm rms}\) implicitly through \(\tau_{\alpha}\).

\subsection{Power spectral density}
\label{subsec4.3:Power_spectral_density}
A complementary way to assess the detectability of the GW signal is to calculate its power spectral density, \(S(f)\), where \(f\) is the Fourier frequency, and compare it with the sensitivity curves of the LIGO detectors.
The power spectral density measures how radiated power is distributed in the frequency domain and helps to guide narrowband searches.
In the stationary regime, i.e.\ \(\tau_{\alpha} \ll \min(t, t')\) and \(C_{\alpha}(t, t') = C_{\alpha}(\zeta)\), we compute the one-sided \(S(f)\) by Fourier transforming equation~\eqref{eqn:K_h} with respect to \(\zeta\), viz.
\begin{align}
    S(f) = 2 \int_{-\infty}^{\infty} \text{d}\zeta \, C_{\alpha}(\zeta) \exp(2\pi i f \zeta) .
    \label{eqn:PSD_definition_stationaryProcess}
\end{align}

\begin{figure}
    \centering
    \includegraphics[width=\columnwidth]{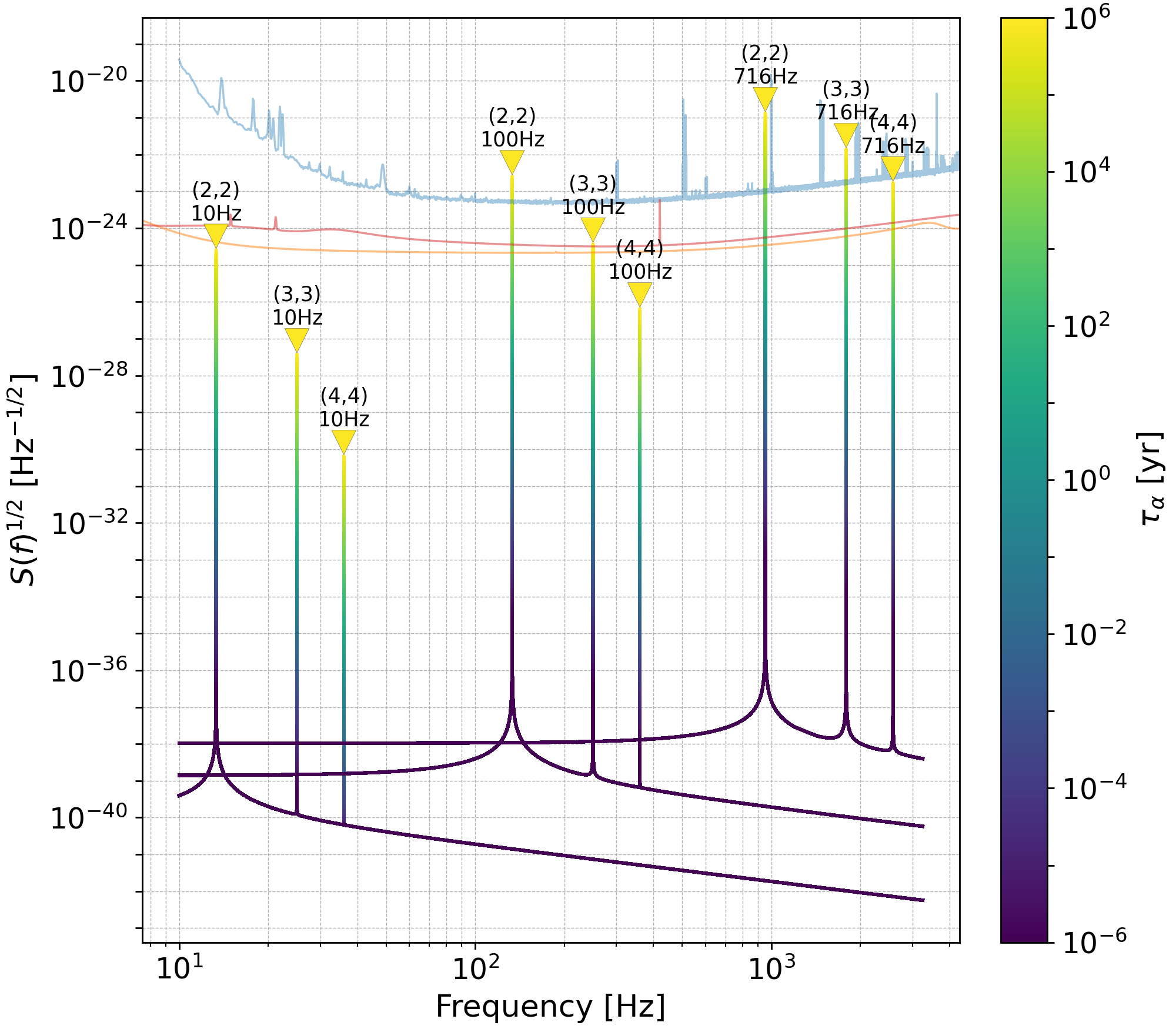}
    \caption{Amplitude spectral density \(S(f)^{1/2}\) versus Fourier frequency \(f\) within the LIGO observing band, for \(l=m=2\), 3, and 4 and \(\nu_{\rm s} = 10 \,\)Hz, 100\,Hz, and 716\,Hz.
    Modes are labelled by \{\((l,m), \nu_{\rm s}\)\}.
    The peaks of the amplitude spectral density are colour-coded by the damping timescale \(\tau_{\alpha} \in [10^{-6}, 10^{6}]\)\,yr. 
    The color gradient along each curve is an illusion; every curve has a single color, but multiple curves overlap indistinguishably except near their peaks.
    The sensitivity curve achieved during the third LIGO observing run (blue curve) \citep{O3SensitivityCurves2021} and projected sensitivity curves for the Einstein Telescope (baseline 10 km, red curve) \citep{HildEtAl2011} and Cosmic Explorer (baseline 40 km, orange curve) \citep{EvansEtAl2021} are also shown for comparison.
    The parameter \(\left| \hat{\bm{v}} \cdot \left[(\bm{x} \times \nabla)Y_{lm}(\hat{\bm{x}})\right]_{\bm{x} = \bm{x}_{\rm s}} \right|\) is set to unity as an upper bound.
    Other parameters: \(R_{*} = 10\,\)km, \(d = 1\,\)kpc, \(\dot{M} = 10^{-8}\,\MsunPerYr\), \(f_{\rm acc} = 1\,\)kHz, and \(|\bm{v}| = 0.4c\).
    }
    \label{fig:ASD}
\end{figure}

Examples of \(S(f)^{1/2}\) in the LIGO observing band are plotted in Figure~\ref{fig:ASD}, for three spin frequencies, viz.\ \(\nu_{\rm s}=10\)Hz, 100Hz, and 716Hz\footnote{
    This is the spin frequency of the fastest spinning neutron star, PSR~J1748-2446ad, known at the time of writing \citep{HesselsEtAl2006}.
    Millisecond pulsars are recycled and have an accretion history.
}, and $r$-modes with \(l=m=2\), 3, and 4.
Each peak is actually $601$ peaks, with \(\tau_{\alpha}\) spanning \(10^{-6} \leq \tau_{\alpha} / (1\,\text{yr}) \leq 10^{6}\), which almost overlap.
The color gradient along each peak, defined by the color bar at the right of the figure, describes how \(S(f)^{1/2}\) near \(f\approx\omega_{\alpha}/(2\pi)\) varies with \(\tau_{\alpha}\) for multiple peaks.
The difference in \(S(f)^{1/2}\) away from the peak for different \(\tau_{\alpha}\) is not visually apparent, due to the broad logarithmic scale.
The damping timescale \(\tau_{\alpha}\) serves as a proxy for the EOS, as discussed in Section~\ref{subsec4.2:Root_mean_square_wave_strain_without_CFS_instabilities}.
We see that the peaks grow narrower and decrease in height as \(l=m\) increases. 
The highest peak reaches \(\max_{f} S(f)^{1/2} \approx 1.3 \times 10^{-21} \text{Hz}^{-1/2}, 1.4 \times 10^{-22} \text{Hz}^{-1/2}\), and \(1.7 \times 10^{-23} \text{Hz}^{-1/2}\) for \(l=m=2\), 3, and 4, respectively, at \(\nu_{\rm s} = 716\)Hz and \(\tau_{\alpha} = 10^{6}\)yr.
The peak height scales \(\propto \nu_{\rm s}^{2} \tau_{\alpha}\), for \(\omega_{\alpha} \tau_{\alpha} \gg 1\).

For modes with \(l=m=2\) or 3, the peaks of \(S(f)^{1/2}\) exceed the LIGO sensitivity (blue curve) in the O3 observing run \citep{O3SensitivityCurves2021} for \(\tau_{\alpha} \gtrsim 10^{5}\)yr.
However, this does not imply automatically, that the GW signal is detectable. 
One needs an integration time \(T_{\rm obs} \gtrsim \tau_{\alpha}\) to detect the peak associated with the mode \(\alpha\).
As one example, for the fastest spinning neutron star at \(\nu_{\rm s} = 716\)Hz and \(d\approx 5.5\)kpc \citep{HesselsEtAl2006,OrtolaniEtAl2007} with \(\tau_{\alpha} = 10^{6}\)yr, one needs \(T_{\rm obs} \gtrsim 10^{3}\)yr to reach the LIGO sensitivity curve.
Therefore, the GW signal from impact-excited $r$-modes is not detectable by the current generation of LIGO interferometers, consistent with Section~\ref{subsec4.2:Root_mean_square_wave_strain_without_CFS_instabilities}.
On the other hand, the observing time requirements are less demanding for the Einstein Telescope \citep{HildEtAl2011} or Cosmic Explorer \citep{EvansEtAl2021}, with \(T_{\text{obs}} \sim 1\)\,yr, for the parameters above.
Signals from low-mass X-ray binaries with \(\tau_{\alpha} \lesssim 10^{5}\)\,yr remain challenging to detect even with the next generation of interferometers.

\section{CFS instability}
\label{sec:5-CFS_instability}
In the preceding sections, we assume that the $r$-modes are CFS-stable.
The assumption is motivated observationally by the nondetection of GWs from $r$-modes, which implies that the dimensionless $r$-mode amplitude satisfies \(\alpha_{r} \lesssim 10^{-9}\) for millisecond pulsars at frequencies inside the nominal instability window \citep{HoEtAl2019,BoztepeEtAl2020}.
The CFS instability is triggered, when \(\Omega\) exceeds a critical value \(\Omega_{\rm CFS}\), leading to \(\tau_{\alpha} < 0\) for \(\Omega > \Omega_{\rm CFS}\).
The nominal instability window is computed, assuming that \(\tau_{\alpha}\) depends only on the gravitational radiation reaction, bulk viscosity from modified URCA reactions, and shear viscosity from neutron-neutron and electron-electron scattering \citep{FriedmanStergioulas2013,GlampedakisGualtieri2018,HoEtAl2019}.
Generalisations of the minimal damping model, which lie outside the scope of this paper, include Ekman layer damping at the crust-core interface \citep{BildstenUshomirsky1999,LevinUshomirsky2001,BondarescuEtAl2007}, mutual friction \citep{HaskellEtAl2009}, and hyperons or strange quarks \citep{NayyarOwen2006,AlfordEtAl2012}.
The reader is directed to Section~9.6.2 of \citet{FriedmanStergioulas2013} and Sections~V.E--V.G of \citet{GlampedakisGualtieri2018} for further details and additional references.
It is instructive to analyse CFS-stable $r$-modes on an equal footing with the stably excited $f$-, $p$-, and $g$-modes, extending the study by \citet{DongMelatos2024}.
The resulting, stochastic GW signal represents quiescent-like emission, which complements the quasimonochromatic emission from CFS-unstable $r$-modes analysed by other authors \citep{Andersson1998,FriedmanMorsink1998,OwenEtAl1998}.

Once the instability switches on, the $r$-mode amplitude grows exponentially. The saturation of the $r$-mode amplitude at a value \(|c_{\alpha, \text{sat}}|\) through nonlinear mode couplings has been analysed in detail in the literature \citep{SchenkEtAl2001,ArrasEtAl2003,BondarescuEtAl2007}.
The instability is quenched, as the gravitational radiation carries away angular momentum, lowering \(\Omega\) below \(\Omega_{\rm CFS}\).

In this section, we briefly discuss how the temporal autocorrelation function \(C_{\alpha}(t, t')\) is modified, when the CFS instability switches on and off.
We consider a single on-off episode, noting that an ongoing cycle of on-off episodes is likely to occur in reality \citep{Levin1999,Heyl2002,ArrasEtAl2003,BondarescuEtAl2007,GlampedakisGualtieri2018}.
For a single on-off episode, we write
\begin{align}
    \tau_{\alpha}^{-1}(t) 
    =& [1 - H(t-t_{\rm on})] \tau_{\alpha, \text{off}}^{-1} 
    + [H(t-t_{\rm on}) - H(t-t_{\rm off})] \tau_{\alpha, \text{on}}^{-1}
    \nonumber \\ &
    + H(t-t_{\rm off}) \tau_{\alpha, \text{off}}^{-1} ,
    \label{eqn:tau_WithOnOff}
\end{align}
where \(t_{\rm on}\) and \(t_{\rm off} > t_{\rm on}\) are the on-off trigger epochs respectively, and \(\tau_{\alpha, \text{on}}\) and \(\tau_{\alpha, \text{off}}\) are the damping timescales when the CFS instability is on and off, respectively.
Equation~\eqref{eqn:tau_WithOnOff} is a simplified model, which relies on two assumptions: (i) \(\tau_{\alpha, \text{off}}\) and \(\tau_{\alpha, \text{on}}\) are constant, with an instantaneous transition between the CFS-stable and the CFS-unstable regime; and (ii) \(\tau_{\alpha, \text{off}}\) is assumed to be the same for \(t < t_{\rm on}\) and \(t > t_{\rm off}\).
The former assumption is justified by treating \(\tau_{\alpha, \text{off}}\) and \(\tau_{\alpha, \text{on}}\) as averaged, effective damping/growth timescales in the CFS-stable and CFS-unstable regimes, respectively.
The latter assumption is likely to be inaccurate immediately after the onset of the CFS-stable regime.
In future work, the calculation will be improved by replacing \(\tau_{\alpha}\) with \(\tau_{\alpha}(t)\) calculated numerically to account for spin and temperature evolution \citep{Levin1999,BondarescuEtAl2007}.

The dominant contribution to \(C_{\alpha}(t, t') \propto K_{\alpha}[\zeta, \min(t, t')]\) is given by the second term in equation~\eqref{eqn:autocorrelation_function_c_alpha_general}, and reads
\begin{align}
    K_{\alpha}(\zeta; t) 
    &\!\approx
    4 f_{\rm acc} \left\langle \cos[\omega_{\alpha} (t - t_{\rm s})] \cos[\omega_{\alpha} (t' - t_{\rm s})] \right\rangle _{\rm s}
    \nonumber \\ & \quad \times 
    \int_{0}^{t} \text{d}t_{\rm s} \, 
    \exp\left[- \int_{t_{\rm s}}^{t} \text{d}s \, \tau_{\alpha}^{-1}(s)\right] 
    \exp\left[- \int_{t_{\rm s}}^{t'} \text{d}s \, \tau_{\alpha}^{-1}(s)\right]
    \nonumber \\ &\!=
    f_{\rm acc} \tau_{\alpha, \text{off}}
    \cos(\omega_{\alpha} \zeta) 
    \exp(-\zeta / \tau_{\alpha, \text{off}})
    \nonumber \\ & \quad \times
    \bigg\{
        1 - \exp\left[
            - 2 (\tilde{t} - \tilde{T}_{\rm CFS})
            - 2 \tilde{T}_{\rm CFS} / \tilde{\tau}_{\alpha, \text{on}}
        \right] 
        \nonumber \\ & \qquad
        + (1 - \tilde{\tau}_{\alpha, \text{on}}) 
        \left[
            \exp\left(
                - 2 \tilde{T}_{\rm CFS} / \tilde{\tau}_{\alpha, \text{on}}
            \right)
            - 1
        \right]
        \nonumber \\ & \qquad \times
        \exp\left[
            - 2 (\tilde{t} - \tilde{t}_{\rm off})
        \right] 
    \bigg\} ,
    \label{eqn:C_2_autocorrelation_CFS}
\end{align}
for \(t' > t > t_{\rm off}\), where a tilde above a temporal quantity indicates, that the quantity is normalised by \(\tau_{\alpha, \text{off}}\). 
The duration of the CFS instability is denoted by \(T_{\rm CFS} = t_{\rm off} - t_{\rm on}\).
Equation~\eqref{eqn:C_2_autocorrelation_CFS} reduces to equation~\eqref{eqn:C_2_autocorrelation} for \(\tau_{\alpha, \text{on}} = \tau_{\alpha, \text{off}}\) or \(T_{\rm CFS} = 0\), i.e. when the CFS instability is not triggered.

Interestingly, equation~\eqref{eqn:C_2_autocorrelation_CFS} can increase or decrease with \(t\) depending on the values of \(\tilde{t}_{\rm off}, \tilde{T}_{\rm CFS}\), and \(\tilde{\tau}_{\alpha, \text{on}}\), whereas equation~\eqref{eqn:C_2_autocorrelation} always increases with \(t\).
From equation~\eqref{eqn:C_2_autocorrelation_CFS}, we find
\begin{align}
    \sgn\left[
        \frac{\partial C_{\alpha}(\zeta; t)}{\partial t}
    \right]
    =\,& 
    \sgn\big\{
    1 - 
    \left(1 - \tilde{\tau}_{\alpha, \text{on}}\right) 
    \left[
        1 - \exp\left(2 \tilde{T}_{\rm CFS} / \tilde{\tau}_{\alpha, \text{on}}\right)
    \right] 
    \nonumber \\ &\times
    \exp\left(
        2\tilde{t}_{\rm on}
    \right)
    \big\} .
    \label{eqn: sign_dCdt_CFS}
\end{align}
In deriving equation~\eqref{eqn: sign_dCdt_CFS}, we assume \(\left|\text{d} \ln{\nu_{\rm s}} / \text{d} t\right| \ll \left|\partial \ln{K_{\alpha}(\zeta; t)} / \partial t\right|\), which is valid throughout the observation span, \(\left|\text{d} \ln(\dot{M} |\bm{v}|) / \text{d} t\right| \ll \left|\partial \ln{K_{\alpha}(\zeta; t)} / \partial t\right|\), and \(\left|\text{d} \ln{f_{\rm acc}} / \text{d} t\right| \ll \left|\partial \ln{K_{\alpha}(\zeta; t)} / \partial t\right|\).
For \(
    \tilde{T}_{\rm CFS} > \tilde{T}_{\rm CFS, crit} 
    = |\tilde{\tau}_{\alpha, \text{on}}| \ln[(1 + |\tilde{\tau}_{\alpha, \text{on}}|) / |\tilde{\tau}_{\alpha, \text{on}}|] / 2
\), one always has \(\partial C_{\alpha}(\zeta; t) / \partial t < 0\), for all \(\tilde{t}_{\rm on} \geq 0\).
Physically, this is because the CFS instability, coexisting with impact-excited $r$-modes, lasts long enough, so that \(C_{\alpha}(\zeta; t > \tilde{t}_{\rm off})\) exceeds the limit \(C_{\alpha, \text{sat}}(\zeta)\) set by the saturated number of clumps during \(\tau_{\alpha}\), regardless of the initial amplitude.
For \(\tilde{T}_{\rm CFS} < \tilde{T}_{\rm CFS, crit}\) on the other hand, one has \(\partial C_{\alpha}(\zeta; t) / \partial t > 0\) for \(\tilde{t}_{\rm on} < \tilde{t}_{\rm on, crit}\), and vice versa.
These regimes, where \(\tilde{T}_{\rm CFS} < \tilde{T}_{\rm CFS, crit}\), correspond to \(C_{\alpha}(\zeta; t > \tilde{t}_{\rm off}) < C_{\alpha, \text{sat}}(\zeta)\), if the initial amplitude is small enough, or \(C_{\alpha}(\zeta; t > \tilde{t}_{\rm off}) > C_{\alpha, \text{sat}}(\zeta)\), if the initial amplitude is high enough.
The critical value of \(\tilde{t}_{\rm on, crit}\) is given by \(
    \tilde{t}_{\rm on, crit} 
    = 
    - \ln\{(1 + |\tilde{\tau}_{\alpha, \text{on}}|) [1 - \exp(-2 \tilde{T}_{\rm CFS} / |\tilde{\tau}_{\alpha, \text{on}}|)]\} / 2
\).

What are the characteristic astrophysical numbers associated with the above regimes?
We estimate \(T_{\rm CFS}\) by the spin-down timescale \(T_{\rm sd} = \Omega / |\dot{\Omega}| \approx 5 |c_{\alpha, \text{sat}}|^{-2} |\tau_{\alpha, \text{on}}|\) \citep{Levin1999,GlampedakisGualtieri2018}. 
For a system that does not reach \(|\tau^{\text{gw}}_{\alpha}| = \tau^{\text{visc}}_{\alpha}\) for \(t > t_{\rm off}\) [i.e.\ a cycle with thermogravitational runaway, followed by saturation], viscous damping is dominated by the shear viscosity, with \(\tau_{22, \text{off}} \approx 1\,(T / 10^{9}\,\text{K})^{2} \,\)yr from electron-electron scattering for temperatures \(\lesssim 10^{9}\,\text{K}\) \citep{FriedmanStergioulas2013}.
One then requires \(|c_{\alpha, \text{sat}}|\) of order unity in order to achieve \(\tilde{T}_{\rm CFS} < \tilde{T}_{\rm CFS, crit}\), for an optimistic value \(|\tau_{\alpha, \text{on}}| = |\tau^{\text{gw}}_{\alpha}| \approx 7 \times 10^{-7} (\nu_{\rm s} / 1 \, {\rm kHz})^{-6}\,\text{yr}\).\footnote{
    See Appendix~\ref{appA3:Growth_timescale} for details about estimating \(\tau^{\text{gw}}_{\alpha}\).
}
Therefore, a neutron star whose CFS instability evolves cyclically almost always has \(\partial C_{\alpha}(\zeta; t) / \partial t < 0\).
On the other hand, if a neutron star evolves towards equilibrium on the $r$-mode stablity curve \citep{BondarescuEtAl2007}, we have \(T_{\rm CFS, crit} \rightarrow \infty\) and \(\partial C_{\alpha}(\zeta; t) / \partial t \propto 2 f_{\rm acc} > 0\).

Table~\ref{table:autocorrelation_CFS_cases} lists the sign of \(\partial C_{\alpha}(\zeta; t>t_{\rm off}) / \partial t\) in three regimes and the possible physical scenarios leading up to the observation span in each regime.
The second order time derivative \(\partial^2 C_{\alpha}(\zeta; t) / \partial t^2 = -(2 / \tau_{\alpha}) \partial C_{\alpha}(\zeta; t) / \partial t\) can be used to discriminate whether the CFS instability is on or off during the observation span, with \(\tau_{\alpha} = \tau_{\alpha, \text{on}} < 0\) and \(\tau_{\alpha} = \tau_{\alpha, \text{off}} > 0\), respectively.

In summary, the slope and concavity of \(C_{\alpha}[\zeta, \min(t, t')]\) can be used to infer observationally the coexistence or otherwise of the CFS instability with impact-excited the $r$-mode oscillations, at least in principle. 
That is, by autocorrelating the wave strain \(h(t)\) detected by an instrument like LIGO, we can infer whether the CFS instability is on or off before or during the observation span, assuming the change in \(\nu_{\rm s}\) is negligible.
We can also infer upper/lower bounds on the duration of the CFS instability or the duration of the uninterrupted accretion episode, compared to the unknown damping timescale \(\tau_{\alpha, \text{off}}\), even if these durations are longer than the observation span itself.

\begin{table}
    \centering
    \begin{tabular}{|c|c|c|}
    \hline
    Regime & \(\partial C_{\alpha}(\zeta; t) / \partial t\) & Physical scenario
    \\ \hline
    \(\tilde{T}_{\rm CFS} > \tilde{T}_{\rm CFS, crit}\) & Negative & Thermogravitational cycle \\
    for all \(\tilde{t}_{\rm on}\) & & 
    \\ \hline
    \(\tilde{T}_{\rm CFS} < \tilde{T}_{\rm CFS, crit}\) & Negative & Indeterminate \\
    \(\tilde{t}_{\rm on} > \tilde{t}_{\rm on, crit}\) & &  
    \\ \hline
    \(\tilde{T}_{\rm CFS} < \tilde{T}_{\rm CFS, crit}\) & Positive & (1) No CFS instability, or \\
    \(\tilde{t}_{\rm on} < \tilde{t}_{\rm on, crit}\) & & (2) \(\tau_{\alpha, \text{off}}\rightarrow \infty\) (equilibrium)
    \\ \hline
    \end{tabular}
    \caption{Sign of \(\partial C_{\alpha}(\zeta; t>t_{\rm off}) / \partial t\) and the associated physical scenario for three regimes defined in terms of the dimensionless parameters \(\tilde{T}_{\rm CFS}\) and \(\tilde{t}_{\rm on}\). 
    The critical values are 
    \(\tilde{T}_{\rm CFS, crit} = |\tilde{\tau}_{\alpha, \text{on}}| \ln[(1 + |\tilde{\tau}_{\alpha, \text{on}}|) / |\tilde{\tau}_{\alpha, \text{on}}|] / 2\) 
    and 
    \(\tilde{t}_{\rm on, crit} = - \ln\{(1 + |\tilde{\tau}_{\alpha, \text{on}}|) [1 - \exp(-2 \tilde{T}_{\rm CFS} / |\tilde{\tau}_{\alpha, \text{on}}|)]\} / 2\).}
    \label{table:autocorrelation_CFS_cases}
\end{table}

\section{Conclusions}
\label{sec:6-conclusions}
The gravitational radiation emitted by $r$-modes that are mechanically excited by the stochastic impact of clumps of matter on an accreting neutron star is studied.
The standard results for the $r$-mode eigenfrequencies and eigenfunctions, in an unmagnetised, slowly-rotating, one-component barotrope are reproduced, assuming Newtonian gravity.
We then calculate analytically, using a Green's function, the response of the $r$-modes to impulsive, top-hat impacts which obey a Poisson point process. 
The stochastic radiated signal is quantified in terms of the instantaneous and root-mean-square wave strain [equations~\eqref{eqn:strain_single_impact} and \eqref{eqn:h_rms_unsaturated} respectively], as well as the power spectral density (Figure~\ref{fig:ASD}).

The instantaneous wave strain excited by a single clump, \(h_0(t; t_{\rm s})\), does not depend strongly on the EOS.
The root-mean-square wave strain satisfies \(
h_{\rm rms} = N_{\rm clumps}^{1/2} h_0(t; t_{\rm s})
\), where \(N_{\rm clumps}\) is the characteristic number of clumps striking the stellar surface, while the $r$-modes oscillate, which depends implicitly on the EOS through the damping timescale \(\tau_{\alpha}\).
For an uninterrupted accretion episode with duration \(\Delta t_{\rm acc} \ll \tau_{\alpha}\), one has \(
h_{\rm rms} \lesssim 10^{-35} (\nu_{\rm s}/ 10\,{\rm Hz})^{2} (\Delta t_{\rm acc}/1\,{\rm yr})^{1/2}
\), with the exact value depending on the impact velocity direction \(\hat{\bm{v}}\) and the impact location \(\bm{x}_{\rm s}\).
This is comparable to the strain from impact-excited $f$-, $p$-, and $g$-modes in a non-rotating star \citep{DongMelatos2024}, and is too weak to be detected by the current generation of LIGO detectors.
The amplitude spectral density peaks at \(\max_f S^{1/2}(f) \sim 10^{-22} \text{Hz}^{-1/2}\) and thereby exceeds nominally the LIGO sensitivity curve in the O3 observing run for \(l=m=2\) $r$-modes with \(\nu_{\rm s} = 100\,\)Hz and \(\tau_{\alpha} \sim 10^{6}\)yr.
The peak amplitude spectral density scales approximately as \(\propto \nu_{\rm s}^{2} \tau_{\alpha}\).
However, the signal is not detectable, as the required integration time, \(T_{\rm obs} \gtrsim \tau_{\alpha} \sim 10^{6}\)yr, is much longer than a typical LIGO observation, consistent with the conclusion drawn from \(h_{\rm rms}\).

The slope and concavity of the temporal autocorrelation function offer an observational route to test for the coexistence of the CFS instability with impact-excited $r$-modes.
Specifically, coexistence strictly before the GW observation implies \(\partial C_{\alpha}(\zeta; t) / \partial t < 0\), and coexistence strictly during the GW observation implies \(\partial C_{\alpha}(\zeta; t) / \partial t > 0\) and \(\partial^2 C_{\alpha}(\zeta; t) / \partial t^2 > 0\).

\section*{Acknowledgements}
This research is supported by the Australian Research Council (ARC) through the Centre of Excellence for Gravitational Wave Discovery (OzGrav) (grant number CE230100016).
We thank the anonymous referee for their constructive feedback.

\section*{Data Availability}

There are no data generated in this paper.



\bibliographystyle{mnras}
\bibliography{GWOANSRmode} 




\appendix

\section{Derivation of r-mode properties}
\label{appA:derviation_r_mode}

In this appendix, we derive in abridged form some key properties of purely axial $r$-modes, which serve as inputs into the GW calculations in Sections~\ref{sec:4-gravitational_radiation} and \ref{sec:5-CFS_instability}.
Eigenfunctions and eigenfrequencies are discussed in Appendix~\ref{appA1:r_mode_frequency_and_eigenfunctions}. Normalisation is discussed in Appendix~\ref{appA2:r_mode_normalisation_constants}.
The gravitational wave growth timescale due to the CFS instability is discussed in Appendix~\ref{appA3:Growth_timescale}.

\subsection{Frequencies and eigenfunctions}
\label{appA1:r_mode_frequency_and_eigenfunctions}
We work in a basis of unit vectors (\(\hat{\bm{r}}, \hat{\bm{\theta}}, \hat{\bm{\phi}}\)) in spherical polar coordinates.
Equation~\eqref{eqn:mode_function_r_mode_general} expressed in component form is given by
\begin{align}
    \bm{\xi}_{\alpha}^{(0)} = \sum_{l=|m|} \frac{U_{l}(r)}{r} \left[ - \frac{1}{\sin \theta} \frac{\partial Y_{lm}}{\partial \phi} \hat{\bm{\theta}} + \frac{\partial Y_{lm}}{\partial \theta} \hat{\bm{\phi}}\right] .
    \label{eqn:mode_function_r_mode_spherical_component}
\end{align}
We write the angular velocity as \(\bm{\Omega} = \Omega \hat{\bf z} = \Omega \cos{\theta} \hat{\bm{r}} - \Omega \sin{\theta} \hat{\bm{\theta}}\), and hence obtain
\begin{align}
    \mathbfss{B} \cdot \bm{\xi}_{\alpha}^{(0)} 
    = -2 \Omega \sum_{l=|m|}^{\infty} \frac{U_{l}}{r} \bigg\{
    \sin{\theta} \frac{\partial Y_{lm}}{\partial \theta} \hat{\bm{r}}
    + \cos{\theta} \frac{\partial Y_{lm}}{\partial \theta} \hat{\bm{\theta}}
    + \frac{\cos{\theta}}{\sin{\theta}} \frac{\partial Y_{lm}}{\partial \phi} \hat{\bm{\phi}}
    \bigg\}.
    \label{eqn:B_dot_xi_r_mode_spherical_coordinates}
\end{align}
Upon substituting \eqref{eqn:mode_function_r_mode_spherical_component} and \eqref{eqn:B_dot_xi_r_mode_spherical_coordinates} into \eqref{eqn:second_order_eigenvalue_problem_2}, 
the \(\hat{\bm{r}}\) component reduces to 
\begin{align}
    \sum_{l=|m|}^{\infty} \frac{U_{l} \, Y_{lm}}{r^2}  \left[2 m \Omega - l(l+1) \omega_{\alpha}^{(1)}\right] \omega_{\alpha}^{(1)} = 0 .
    \label{eqn:r_mode_frequency_equation}
\end{align}
After applying the orthogonality condition of the spherical harmonics, equation~\eqref{eqn:r_mode_frequency_equation} implies
\begin{align}
    \omega_{\alpha}^{(1)} = \frac{2 m \Omega}{l(l + 1)} ,
    \label{eqn:r_mode_frequency_appendix}
\end{align}
for an individual \(l \geq |m|\), and \(U_{l} = 0\) otherwise.
The \(\hat{\bm{\theta}}\) component reduces to 
\begin{align}
    &\sum_{l=|m|}^{\infty} 
    \left[\left(1 - \frac{l \omega_{\alpha}^{(1)}}{2m\Omega}\right) U_{l}' - \frac{l U_{l}}{r}\right] Q_{l+1 \, m} Y_{l+1 \, m}
    \nonumber \\ &\qquad
    + \left\{\left[1 + \frac{(l+1) \omega_{\alpha}^{(1)}}{2m\Omega}\right] U_{l}' + \frac{(l+1) U_{l}}{r}\right\} Q_{l \, m} Y_{l-1 \, m} 
    = 0,
    \label{eqn:r_mode_function_eqn_theta_component}
\end{align}
where the prime denotes a derivative with respect to \(r\), and we exploit the recursion relations
\begin{align}
    \sin{\theta} \frac{\partial Y_{lm}}{\partial \theta} &= l Q_{l+1 \, m} Y_{l+1 \, m} - (l+1) Q_{l \, m} Y_{l-1 \, m}, \\
    \cos{\theta} Y_{lm} &= Q_{l+1 \, m} Y_{l+1 \, m} + Q_{l \, m} Y_{l-1 \, m}, 
\end{align}
with
\begin{align}
    Q_{l \, m} &= \left[\frac{(l - m)(l + m)}{(2l + 1)(2l - 1)}\right]^{1/2}.
\end{align}
Equation~\eqref{eqn:r_mode_function_eqn_theta_component} 
agrees with equation (35)
in \citet{LockitchFriedman1999} for \(U_{l}(r) \neq 0\).
Upon applying \eqref{eqn:r_mode_frequency_appendix} and the orthogonality condition of the spherical harmonics to \eqref{eqn:r_mode_function_eqn_theta_component} 
, we arrive at two equations for the $r$-mode eigenfunctions \citep{GittinsAndersson2023},
\begin{align}
    0 &= Q_{l+1 \, m} \left[U_{l}' - \frac{(l+1) U_{l}}{r}\right] , 
    \label{eqn:r_mode_eigenfunction_equation1}
    \\ 
    0 &= Q_{l \, m} \left(U_{l}' + \frac{l U_{l}}{r}\right) .
    \label{eqn:r_mode_eigenfunction_equation2}
\end{align} 
In order to simultaneously satisfy equations~\eqref{eqn:r_mode_eigenfunction_equation1} and \eqref{eqn:r_mode_eigenfunction_equation2}, we must have \(l = |m|\), and
\begin{align}
    U_{|m|} \propto r^{|m| + 1} .
\end{align}
Therefore, we have shown that one has \(U_{l} \propto r^{l + 1}\) for \(l = |m|\), and \(U_{l} = 0\) otherwise \citep{FriedmanStergioulas2013}.

\subsection{Normalisation}
\label{appA2:r_mode_normalisation_constants}
In this subsection, we derive the leading-order normalisation constants \(\mathcal{N}_{\alpha}\), in order to quantify the gravitational response of the $r$-modes to the mechanical impact of accreted matter in Section~\ref{sec:4-gravitational_radiation}.

We introduce the slow-rotation expansion of the equilibrium density \(\rho_0\) in the inner product \eqref{eqn:inner product} \citep{SchenkEtAl2001} and express it as
\begin{align}
    \langle \bm{\xi}_{\alpha}, \bm{\xi}_{\alpha} \rangle 
    &= \int \text{d}^3 x \, \left[\rho_0^{(0)} + \rho_0^{(2)} + \Order{\Omega^4}\right]
    \bm{\xi}_{\alpha}^{*} \cdot \bm{\xi}_{\alpha} ,
    \\
    &= \langle \bm{\xi}_{\alpha}, \bm{\xi}_{\alpha} \rangle^{(0)} + \langle \bm{\xi}_{\alpha}, \bm{\xi}_{\alpha} \rangle^{(2)} + \Order{\Omega^4} .
\end{align}
The angular frequencies of $r$-modes is \(\sim \Order{\Omega}\). 
Hence, to leading order, we obtain
\begin{align}
    \mathcal{N}_{\alpha}^{(1)} 
    = 2 \omega_{\alpha}^{(1)} \kappa_{\alpha}^{(0)} \langle \bm{\xi}_{\alpha}^{(0)}, \bm{\xi}_{\alpha}^{(0)} \rangle^{(0)} ,
\end{align}
with
\begin{align}
    \kappa_{\alpha}^{(0)} = 1 - \frac{1}{2 \omega_{\alpha}^{(1)}} \frac{\langle \bm{\xi}_{\alpha}^{(0)}, i \mathbfss{B}^{(1)} \cdot \bm{\xi}_{\alpha}^{(0)} \rangle^{(0)}}{\langle \bm{\xi}_{\alpha}^{(0)}, \bm{\xi}_{\alpha}^{(0)} \rangle^{(0)}} .
    \label{eqn:kappa^0_definition}
\end{align}
Upon substituting \eqref{eqn:mode_function_r_mode_spherical_component} and \eqref{eqn:B_dot_xi_r_mode_spherical_coordinates} into \eqref{eqn:inner product}, we obtain
\begin{align}
    &\langle \bm{\xi}_{\alpha}^{(0)}, i \mathbfss{B}^{(1)} \cdot \bm{\xi}_{\alpha}^{(0)} \rangle^{(0)}
    \nonumber \\ 
    &= 2 m \Omega \int \text{d}r \, \rho_0 U_{l}^2 
    \int \text{d}\phi \text{d}(\cos{\theta}) \, \cot{\theta} (Y_{lm}^{*} \partial_{\theta} Y_{lm} + \text{c.c.}) ,
    \label{eqn:inner_product_B_dot_xi_subbed}
    \\
    &= 2 m \Omega \int \text{d}r \, \rho_0 U_{l}^2 
    \int \text{d}\phi \text{d}(\cos{\theta}) \, \cot{\theta} \, 
    \partial_{\theta}|Y_{lm}|^2 ,
    \label{eqn:inner_product_B_dot_xi_subbed_|Y|^2}
    \\
    &= 2 m \Omega \int \text{d}r \, \rho_0 U_{l}^2
    \int \text{d}\phi \text{d}\theta \, 
    \left[
        \frac{\partial}{\partial \theta} \left(\cos{\theta} |Y_{lm}|^2\right)
        + \sin{\theta} |Y_{lm}|^2
    \right]
    \label{eqn:inner_product_B_dot_xi_subbed_byparted}
    \\
    &= 2 m \Omega \int \text{d}r \, \rho_0 U_{l}^2
    \left[- (2l + 1) \delta_{m0} + 1\right] ,
    \label{eqn:inner_product_B_dot_xi_subbed_final}
    \\
    &= \omega_{\alpha}^{(1)} \langle \bm{\xi}_{\alpha}^{(0)}, \bm{\xi}_{\alpha}^{(0)} \rangle^{(0)} .
    \label{eqn:xi_iBxi^0}
\end{align}
To pass from \eqref{eqn:inner_product_B_dot_xi_subbed_|Y|^2} to \eqref{eqn:inner_product_B_dot_xi_subbed_byparted}, we integrate by parts.
We then use \eqref{eqn:r_mode_frequency_appendix} and
\begin{align}
    \int \text{d}\phi \text{d}(\cos{\theta}) \, |\bm{r} \times \nabla Y_{lm}|^2 = l(l+1)
\end{align} to convert \eqref{eqn:inner_product_B_dot_xi_subbed_final} into \eqref{eqn:xi_iBxi^0}.
Simplifying \eqref{eqn:kappa^0_definition} using \eqref{eqn:xi_iBxi^0}, we obtain \(\kappa_{\alpha}^{(0)} = 1/2\).

\subsection{Growth timescale of the CFS instability}
\label{appA3:Growth_timescale}
In this subsection, we briefly review the theory of the CFS instability of $r$-modes to support Section~\ref{sec:5-CFS_instability}, where the effect of the CFS instability on GW signals from clump-excited $r$-modes is discussed.

The energy of the mode in the rotating frame is given by
\begin{align}
    E_{\alpha, \text{rot}} = \frac{1}{2} \langle{\delta \bm{u}_{\!\alpha}, \delta \bm{u}_{\!\alpha}}\rangle ,
    \label{eqn:r_mode_canonical_energy_rotating_frame}
\end{align}
where angular brackets denote an inner product defined in Section~\ref{subsec2.4:Orthogonality}.
The power radiated by a normalised $r$-mode due to gravitational radiation reaction in the inertial frame is \citep{Thorne1980}
\begin{align}
    \dot{E}_{\alpha, \text{i}}
    = - \frac{G}{32\pi c^{2l+3}} \left|\frac{\text{d}^{l+1} \delta S_{lm}}{\text{d} t^{l+1}}\right|^2 .
    \label{eqn:energy_loss_rate_GW_inertial_frame}
\end{align}
The rotating frame power \(\dot{E}_{\alpha, \text{r}}\) is related to \(\dot{E}_{\alpha, \text{i}}\) via the gravitational radiation reaction torque \(\dot{J}_{\alpha}\), viz.
\begin{align}
    \dot{E}_{\alpha, \text{r}}
    &= \dot{E}_{\alpha, \text{i}} - \Omega \dot{J}_{\alpha}
    \\
    &= \frac{\omega_{\alpha}}{\omega_{\alpha, \text{i}}} \, \dot{E}^{\rm i}_{\alpha} ,
    \label{eqn:energy_loss_rate_GW_rotating_frame}
\end{align}
where we use \(\dot{J}_{\alpha} = - (m / \omega_{\alpha, \text{i}}) \dot{E}_{\alpha, \text{i}}\) and \(\omega_{\alpha, \text{i}} = \omega_{\alpha} - m\Omega\).
We estimate the growth timescale due to gravitational radiation reaction a posteriori by \citep{OwenEtAl1998,FriedmanStergioulas2013}
\begin{align}
    \tau^{\text{gw}}_{\alpha}
    = - \frac{2 \, E_{\alpha, \text{r}}}{\dot{E}_{\alpha, \text{r}}} .
    \label{eqn:tau_GR_definition}
\end{align}
When \(\omega_{\alpha, \text{i}}\) and \(\omega_{\alpha}\) have opposite signs, one obtains \(\dot{E}_{\alpha, \text{r}} > 0\) and \(\tau^{\text{gw}}_{\alpha} < 0\), indicating that such modes are unstable.
This is known as the CFS instability \citep{FriedmanSchutz1978}.

Upon combining equations~\eqref{eqn:omega_r_mode}--\eqref{eqn:r-mode_velocity_perturbation}, \eqref{eqn:deltaS_lm_form2}, and \eqref{eqn:r_mode_canonical_energy_rotating_frame}--\eqref{eqn:tau_GR_definition}, we arrive at
\begin{align}
    \tau^{\text{gw}}_{\alpha}
    = - \frac{c^{2l + 3}}{128\pi G} \frac{l(l + 1)^3 [(2l + 1)!!]^2}{(l - 1)^{2l}} 
        \left(\frac{l+1}{l+2}\right)^{2l+2} \frac{\Omega^{-2l - 2} R^{-2l+2}}{\langle{\bm{\xi}_{\alpha}^{(0)}, \bm{\xi}_{\alpha}^{(0)}}\rangle} .
    \label{eqn: tau_GR}
\end{align}
For $r$-modes with \(l=m=2\) in a uniform density neutron star, equation~\eqref{eqn: tau_GR} reduces to \citep{FriedmanStergioulas2013} 
\begin{align}
    \left|\tau^{\text{gw}}_{\alpha}\right|
    = 22
    \left(\frac{\nu_{\rm s}}{1 \, \text{kHz}}\right)^{-6}
    \left(\frac{M}{1.4 \, \Msun}\right)^{-1}
    \left(\frac{R_{*}}{10 \, \text{km}}\right)^{-4}
    \text{s} .
\end{align}


\bsp	
\label{lastpage}
\end{document}